\documentclass[letterpaper,twocolumn,10pt]{article}
\usepackage{usenix2024_SOUPS}
\usepackage{usenix}
\usepackage{epsfig}
\usepackage{endnotes}
\usepackage{helvet}
\usepackage{enumerate}
\usepackage{array}
\usepackage{tabularx}
\usepackage{xcolor}
\usepackage{colortbl}
\usepackage{tcolorbox}
\usepackage{microtype}
\usepackage{xspace}
\usepackage{adjustbox}
\usepackage{amsmath}
\usepackage{hyperref}
\usepackage{booktabs}
\usepackage{subcaption}
\usepackage{comment}
\usepackage{multirow}
\usepackage{amssymb}
\usepackage[shortlabels]{enumitem}

\usepackage{tikz}
\usepackage{amsmath}
\usepackage{titlesec}

\usepackage{filecontents}
\def\confversion{0}

\usepackage[absolute]{textpos}
\newcommand{\headingg}[1]{\noindent\textbf{#1}}
\newcommand{\heading}[1]{\vspace*{6pt}\noindent\textbf{#1}}

\renewenvironment{quote}
               {\list{}{\topmargin=0.5em\leftmargin=0.5em\rightmargin=0.5em}
                \item\relax}
               {\endlist}

\newcommand{\aigncii}{AIG-NCII\xspace}
\newcommand{\gii}{GII\xspace}

\begin{document}

\ifnum\confversion=0
\begin{textblock}{175}(15,10)
  \noindent A shortened version of this paper appears in the \textit{Proceedings of the 20th Symposium on Usable Privacy and Security (SOUPS)}, August 2024. This is the extended version.
\end{textblock}
\fi

\date{}

\title{\Large \bf``Violation of my body:''\\
Perceptions of AI-generated non-consensual (intimate) imagery}

\author{
{\rm Natalie Grace Brigham}\\
University of Washington
\and
{\rm Miranda Wei}\\
University of Washington
\and
{\rm Tadayoshi Kohno}\\
University of Washington
\and
{\rm Elissa M. Redmiles}\\
Georgetown University
}

\maketitle

\ifnum\confversion=1
\thecopyright
\fi

\begin{abstract}
AI technology has enabled the creation of deepfakes: hyper-realistic synthetic media. We surveyed 315 individuals in the U.S. on their views regarding the hypothetical non-consensual creation of deepfakes depicting them, including deepfakes portraying sexual acts. Respondents indicated strong opposition to creating and, even more so, sharing non-consensually created synthetic content, especially if that content depicts a sexual act. However, seeking out such content appeared more acceptable to some respondents. Attitudes around acceptability varied further based on the hypothetical creator’s relationship to the participant, the respondent’s gender and their attitudes towards sexual consent. This study provides initial insight into public perspectives of a growing threat and highlights the need for further research to inform social norms as well as ongoing policy conversations and technical developments in generative AI.
\end{abstract}

\section{Introduction}
Technological advancements in artificial intelligence (AI) have enabled the creation of hyper-realistic synthetic media known as ``deepfakes.'' This term, a portmanteau of ``deep learning'' and ``fake,'' refers to synthetic image, audio, or video representations of individuals that has been automatically generated using machine learning~\cite{WesterlundDeepfakeReview, FairdDeepfakeReport, KietzmannDeepfakeReview}. Deepfakes encompass many forms of media synthesis, including voice-swapping, text-to-speech, face-swapping, face-morphing, full-body puppetry, and lip syncing~\cite{KietzmannDeepfakeReview}. Moreover, recent progress in generative AI has enabled the creation of deepfakes using only text prompts, rather than requiring a data set of training images depicting the target individual~\cite{PumarolaGeneration, WilliamsGeneration, 404MediaTaylorSwift}. While deepfake technology has potentially benevolent applications in creativity, accessibility, and entertainment~\cite{DanryGoodDeepfake, FairdDeepfakeReport, WesterlundDeepfakeReview, EtienneGoodDeepfakes, CaporussoGoodDeepfakes}, it has also been used to spread disinformation, commit fraud (e.g., phishing), and non-consensually generate intimate imagery~\cite{ChesneyPoliticalDeepfakes, deRancourtPhishing, AjderDeeptrace}.\footnote{Intimate imagery refers to ``images and videos of people who are naked, showing their genitals, engaging in sexual activity or poses, or wearing underwear in compromising positions''~\cite{StopNCII}.} The latter has commonly been termed ``deepfake pornography,'' but following evolving terminology around image-based sexual abuse~\cite{McGlynnRevengePorn}, we refer to it in this paper AI-generated non-consensual intimate imagery (\aigncii).\footnote{\aigncii is our preferred term because it emphasizes the non-consensual nature of the images and is more widely applicable to the range of technologies that can be used to create such images.}

Current technical research around deepfakes has predominantly focused on developing generative AI systems capable of synthesizing such content, including face-swapping~\cite{NirkinFaceSwap, XuFaceSwap} and text-to-video systems~\cite{WuTextToVideo, HongTextToVideo, SingerTextToVideo}, detection methods~\cite{ZhaoDetection, CaldelliDetection, DolhanskyDetection}, as well as strategies to disrupt their generation~\cite{RuizDisruption}. However, research on attitudes of the general public towards deepfakes is far more nascent.  
A large body of literature and theory in information systems and HCI has underscored the importance technology acceptance --- by individuals and by society --- on technology use (and misuse)~\cite{marangunic2015technology,irani2010postcolonial}. Thus, this research seeks to bridge the gap between the technically possible (e.g., the academic research cited above) and the public acceptance of different uses of  the technology. As computer security and privacy researchers, we are particularly interested in adversarial contexts, e.g., the generation of \aigncii.
Hence, we ask:
\textbf{What are people’s attitudes toward the hypothetical non-consensual creation, sharing, and/or seeking out of deepfakes depicting them?} Decomposing this question, we ask specifically:
\begin{enumerate}[leftmargin=1cm, label=\textbf{RQ\arabic*:}, itemsep=0pt]
\item How do attitudes differ depending on what is depicted: \aigncii vs.\ non-consensually created content depicting \textit{non-sexual} acts?
\item How do these attitudes differ depending on contextual factors: who is creating the media and for what purpose?
\item How do attitudes related to sexual (a) consent and (b) content influence these attitudes?
\item How does gender influence these attitudes?
\end{enumerate}

\ifnum\confversion=0
\newpage
\fi
To answer these questions, we conducted a vignette-based survey of 315 individuals to assess attitudes towards different situations involving non-consensual synthetic media. This research elucidates contextual and individual factors that shape public acceptance of generative AI technology being used to construct deepfakes in addition to broader trends in attitudes and rationales. Through this work, we aim to inform future discourse regarding deepfakes, specifically \aigncii, in public, technical, legal, and policy spheres.
\section{Background \& Related Work}
In 2017, a user named ``deepfakes'' posted synthetic videos of celebrities in sexual acts to Reddit~\cite{FairdDeepfakeReport, KuglerDeepfakeAttitudes, KietzmannDeepfakeReview}. Over 90,000 users subsequently joined an r/deepfake subreddit for creating and sharing similar content, drawing significant public attention before being banned by Reddit as ``involuntary pornography''~\cite{RedditPornPolicy}.
Online communities catalyzed the popular use of the term ``deepfake''~\cite{FairdDeepfakeReport, KuglerDeepfakeAttitudes, KietzmannDeepfakeReview}, and despite bans on mainstream social media platforms,
\aigncii continues to be produced and circulated on dedicated forums~\cite{AjderDeeptrace, TimmermanDeepfakeCommunity}. 

\heading{Image-Based Sexual Abuse (IBSA).}
\aigncii is one form of IBSA: the non-consensual creation, distribution, or threats made with intimate images~\cite{McGlynnIBSA, McGlynnIBSAHarms, RuvalcabaIBSA}.
Victim-survivors of IBSA often experience severe health consequences, 
such as post-traumatic stress disorder, anxiety, depression, and greater somatic burdens~\cite{BatesIBSAHarms, EatonIBSAReport, HuberIBSAHarms, RuvalcabaIBSA}. IBSA harms are also social, e.g., isolation, lowered self-esteem, trust issues, and unhealthy coping mechanisms~\cite{BatesIBSAHarms, McGlynnIBSAHarms}. 
Victim-blaming attitudes are prevalent when seeking support or justice after IBSA~\cite{FlynnVictimBlaming}, and obstruct help-seeking~\cite{MennickeDisclosure, OrchowskiDisclosure}.
IBSA falls under a broader umbrella of technology-facilitated gender-based violence~\cite{DunnTFGBV}.
As with other gender-based violence, victim-survivors of IBSA are predominantly, though not exclusively, women~\cite{AjderDeeptrace, DunnDeepfakeVictims}.

IBSA and \aigncii are growing global issues~\cite{FlynnIBSADeepfakes}. Policy on IBSA is sparse in most countries~\cite{ExecutiveOrder, williams2023legal}; in the US specifically, legal scholars have called for legislation to sufficiently address its harms~\cite{CitronDeepfakeCrim, GiesekeDeepfakeCrim, DelfinoDeepfakeCrim}. Understanding public attitudes about synthetic media, specifically \aigncii, can inform better policies on this emergent form of IBSA.

\heading{Public attitudes about \aigncii.}
Early research found significant public concern about non-sexual deepfake creation and dissemination~\cite{GottfriedPewAttitudes}, but less if created for entertainment, humor, or with consent and traceability~\cite{LiDeepfakeAttitudes, NapshinAttitudes}.

Regarding \aigncii, i.e., sexual deepfakes, prior work has primarily focused on attitudes around criminality and perceived harm to victim-survivors~\cite{KuglerDeepfakeAttitudes,FidoCelebrity,UmbachAttitudes}. 
Kugler and Pace found that individuals in the UK perceived significant harms from and strongly favored criminalization of sexual and non-sexual deepfakes~\cite{KuglerDeepfakeAttitudes}.
Further, videos being labeled as fake did reduce the perceived harm of non-sexual deepfakes, but did not for \aigncii~\cite{KuglerDeepfakeAttitudes}.
Fido et al. study \aigncii while varying the identity of the target, finding that deepfakes of celebrities were perceived as less criminal and less harmful, especially for celebrities who are men~\cite{FidoCelebrity}.
This work also found that creation of deepfakes for personal sexual gratification was viewed as less harmful and criminal than sharing.
Finally, in Umbach et al.'s study across ten countries, awareness of \aigncii was low overall, but surveyed individuals believed victims had a right to be upset~\cite{UmbachAttitudes}.
Men in this study also reported more perpetration and victimization.

We combine elements from prior work on non-sexual deepfakes and \aigncii to systematically study \textit{acceptance} (vs. criminality or harm) of the use of generative AI technology to create different types of deepfakes. Specifically, we extend \cite{KuglerDeepfakeAttitudes} to compare \aigncii with not-exclusively-harmful deepfake actions (RQ1): saying something -- which is ambiguous regarding sexuality or harmfulness -- and playing a sport -- ostensibly, a neutral action. We make these comparisons across five disambiguated actions involving deepfakes: creating, private sharing, public sharing, resharing, and seeking out. Additionally, we explore the role of contextual factors (RQ2) such as \textit{intent} of the creator; a factor not explored in prior work on \aigncii despite the fact that intent is a factor in existing laws that can be applied to deepfakes and image-based sexual abuse~\cite{CitronSexualPrivacy} and the fact that prior work on non-sexual deepfakes finds that intent affects the general public's attitudes toward acceptability~\cite{LiDeepfakeAttitudes, NapshinAttitudes}. As a second contextual factor, we further explore the relationship between the creator and subject; we explore the role of intimate partnership while prior work explored, and found relevant, celebrity status~\cite{FidoCelebrity}. We further explore the impact of individual factors on these attitudes. We select individual factors found relevant in prior work on offline sexual abuse such as sexual consent attitudes~\cite{HumphreysSCSR} but which have been unexplored in the context of deepfakes and AIG-NCII (RQ3); as well as individual factors found relevant in prior work on AIG-NCII criminality perceptions such as gender~\cite{UmbachAttitudes} (RQ4). 

Finally, as noted by Fido et al.~\cite{FidoCelebrity}, prior work lacks qualitative exploration of \textit{why} respondents held particular opinions. 
In our work, we collect and analyze qualitative data on attitudes toward the acceptability of creating AIG-NCII and other synthetic media.

\heading{Deepfake community attitudes.} 
Research has examined pro-deepfake views among Reddit users~\cite{GamageDeepfakesReddit} and on MrDeepFakes~\cite{TimmermanDeepfakeCommunity}, as well as positive attitudes but misuse concerns in a deepfake tool's open-source community~\cite{WidderDeepfakesAttitude}.
\section{Methodology}
\label{sec:methodology}

We conducted a survey of 315 U.S. Prolific respondents 
\ifnum\confversion=1
(survey instrument provided in the extended arXiv version of this paper~\cite{full}).
\else
(full survey instrument provided in Appendix~\ref{ap:full-survey-instrument}).
\fi
Our Institutional Review Board (IRB) found our study to be exempt and we followed the ethical considerations as described in Section~\ref{sec:ethics}.

\subsection{Survey structure}
\label{sec:methodology:survey-structure}

\headingg{Consent.} 
The survey began with a description of generative AI and its capacity to generate realistic-looking but fake media. We chose to avoid using ``deepfake'' given potential priming effects (e.g., about political disinformation). Respondents then were told survey structure and asked to consent.

\heading{Vignettes.} \label{sec:methodology:survey-structure:vignette}
We used vignettes---short descriptions of hypothetical scenarios---to solicit respondents' attitudes about \aigncii. Vignettes are common in security and privacy studies to elicit reactions~\cite{EmamiSmartCities, EmamiIoT, MartinVignette} and can approximate real-world behaviors~\cite{HainmullerVignette}. Drawing on the theory of contextual integrity~\cite{NissenbaumPrivacy}, each vignette described generative AI being used to create a video of the respondent without their knowledge, varying three factors:

\begin{enumerate}[label=(\arabic*), nosep]
    \item \textsf{action} varies sexual explicitness, from unambiguously sexual behavior (`performing a sexual act') to non-sexual (`playing a sport') to ambiguous (`saying something'). This factor corresponds to RQ1.
    \item \textsf{creator} varies the relationship between the media maker and participant, either `an intimate partner' or `a stranger.' This corresponds to RQ2 and complements prior work~\cite{KuglerDeepfakeAttitudes, FidoCelebrity} exploring other relationships (e.g., of a celebrity).
    \item \textsf{intent} varies the \textsf{creator}'s motivation, representing motivations reported by prior work~\cite{FairdDeepfakeReport, WesterlundDeepfakeReview}: `harming you,' `entertainment,' and `sexual pleasure,' also corresponding to RQ2.
\end{enumerate}

One such vignette reads: ``Imagine that an intimate partner uses generative AI to create a synthetic video of you playing a sport for the purpose of entertainment. Assume that you are unaware of the video's creation and existence.''
We employed a 2 (\textsf{creator}) × 3 (\textsf{action}) × 3 (\textsf{intent}) full-factorial design to construct 18 vignettes (see Table \ref{tab:fullVignetteTable}).
The six vignettes where \textsf{action} was `performing a sexual act' constitute cases of \aigncii.
Other vignettes, such as V8, are not necessarily \aigncii but may still be sensitive.
Each respondent was randomly assigned three vignettes to mitigate survey fatigue~\cite{PorterSurveyFatigue}. For each vignette, respondents rated the acceptability on a 5-point Likert scale from ``Totally unacceptable'' to ``Totally acceptable''; for ratings other than ``Neutral'', they also wrote a short open-ended rationale about their choice.

Prior work found initial evidence~\cite{FidoCelebrity, KuglerDeepfakeAttitudes, UmbachAttitudes} or hypothesized~\cite{OhmanPervertsDillema, StoryEthics} that acceptability may vary across behaviors. Thus, we assess acceptability for five \aigncii behaviors: 
\begin{enumerate}[label=(\arabic*), nosep]
    \item \textsf{creation} of the video 
    \item \textsf{private\_sharing} by the \textsf{creator}, e.g., in a group chat
\end{enumerate}

\ifnum\confversion=1
\setlength{\textfloatsep}{4pt}
\else
\setlength{\textfloatsep}{8pt}
\fi

\begin{table}[t]
\centering
\resizebox{\columnwidth}{!}{%
\footnotesize
\begin{tabular}{lllll}
\toprule
\textbf{ID} & \textsf{\textbf{creator}} & \textsf{\textbf{action}} & \textsf{\textbf{intent}}  \\
\midrule \rowcolor{yellow!30}
V1 & an \textit{intimate partner} & performing a \textit{sexual act} & \textit{entertainment} \\
\midrule \rowcolor{yellow!30}
V2 & an \textit{intimate partner} & performing a \textit{sexual act} & \textit{harm}ing you \\
\midrule \rowcolor{yellow!30}
V3 & an \textit{intimate partner} & performing a \textit{sexual act} & \textit{sexual pleasure} \\
\midrule
V4 & an \textit{intimate partner} & playing a \textit{sport} & \textit{entertainment} \\
\midrule
V5 & an \textit{intimate partner} & playing a \textit{sport} & \textit{harm}ing you  \\
\midrule
V6 & an \textit{intimate partner} & playing a \textit{sport} & \textit{sexual pleasure}  \\
\midrule
V7 & an \textit{intimate partner} & \textit{saying something} & \textit{entertainment}  \\
\midrule
V8 & an \textit{intimate partner} & \textit{saying something} & \textit{harm}ing you \\
\midrule
V9 & an \textit{intimate partner} & \textit{saying something} & \textit{sexual pleasure}  \\
\midrule \rowcolor{yellow!30}
V10 & a \textit{stranger} & performing a \textit{sexual act} & \textit{entertainment}  \\
\midrule \rowcolor{yellow!30}
V11 & a \textit{stranger} & performing a \textit{sexual act} & \textit{harm}ing you  \\
\midrule \rowcolor{yellow!30}
V12 & a \textit{stranger} & performing a \textit{sexual act} & \textit{sexual pleasure}  \\
\midrule
V13 & a \textit{stranger} & playing a \textit{sport} & \textit{entertainment} \\
\midrule
V14 & a \textit{stranger} & playing a \textit{sport} & \textit{harm}ing you  \\
\midrule
V15 & a \textit{stranger} & playing a \textit{sport} & \textit{sexual pleasure}  \\
\midrule
V16 & a \textit{stranger} & \textit{saying something} & \textit{entertainment}  \\
\midrule
V17 & a \textit{stranger} & \textit{saying something} & \textit{harm}ing you &  \\
\midrule
V18 & a \textit{stranger} & \textit{saying something} & \textit{sexual pleasure}  \\
\bottomrule
\end{tabular}}
\caption{The ID and contextual details of \textsf{creator}, \textsf{action}, and \textsf{intent} of each vignette. The italicized portions of the contextual details are the shorthand descriptions of the vignettes used in the paper text, e.g., V1 - intimate partner/sexual act/entertainment. The highlighted vignettes are \aigncii.} 
\label{tab:fullVignetteTable}
\end{table}

\begin{enumerate}[label=(\arabic*), nosep]
    \setcounter{enumi}{2}
    \item \textsf{public\_sharing} by the \textsf{creator}, e.g., posting it on Reddit
    \item \textsf{resharing}, publicly, by someone who received the video from the \textsf{creator}
    \item \textsf{seeking\_out} by someone with whom it was not shared, e.g., searching online by a description of the video
\end{enumerate}

\heading{Sexual Consent Scale-Revised.} To answer RQ3a about the role of attitudes towards sexual consent, we use two validated subscales from the Sexual Consent Scale-Revised (SCS-R)~\cite{HumphreysSCSR} 
\ifnum\confversion=1
(included in the extended arXiv version~\cite{full})
\else
(Appendix~\ref{ap:full-survey-instrument})
\fi
: SCS-R2 measures attitudes toward establishing consent, and SCS-R4 measures agreement with sexual consent norms based on relationship status and sexual activity. These subscales were selected over others from the SCS-R as our focus was on respondents' attitudes rather than self-reported behaviors.

\heading{Genuine Intimate Imagery (\gii) and NDII Attitudes.} \label{sec:methodology:GII-NDII} To answer RQ3b about attitudes towards sexual content, we assessed attitudes on intimate media creation in intimate relationships.
Paralleling the vignettes, we also asked about four scenarios involving non-consensual distribution of intimate images (NDII): (1) private sharing and (2) public sharing by the intended recipient, as well as (3) public sharing and (4) seeking out by someone who was \textit{not} the intended recipient.

\heading{Demographics.} The survey concluded with demographic questions, including gender (RQ4).

\subsection{Respondents}
We used power analysis to determine the required number of respondents for constructing our regression models with the ability to observe small-to-medium effects.
We recruited 335 Prolific respondents who were over 18, lived in the US, and had over 95\% approval on Prolific. 
20 respondents who did not pass a Pew attention check question~\cite{PewAttention} or provided incoherent open-ended responses were excluded.
The survey took an average of 15 minutes to complete. We compensated respondents \$3, which we calculated based on our average pilot test length (12 minutes) and a rate of \$15/
hour. 156 respondents were women, 150 were men, 6 were non-binary, 2 were agender, and 1 preferred not to say. Further demographic information is presented in Appendix~\ref{ap:demographics}.

\subsection{Data analysis}
\label{sec:methodology:data-analysis}

\headingg{Quantitative analysis.} 
\label{sec:methodology:data-analysis:quant}
Given that the dependent variable was a categorical Likert scale measuring acceptability judgments, and we aimed to include both fixed and random effects as independent variables, we analyzed respondents' attitudes using cumulative link mixed models (CLMMs). 
We built five CLMMs, one for each of the dependent variables concerning the synthetic video described in the vignettes, listed above. 
Each model included the same six independent variables. The first three were the vignette factors (\textsf{creator}, \textsf{action}, \textsf{intent}) (RQ1 \& RQ2). 
For RQ3a, we included participant scores on the two SCS-R subscales. To evaluate potential co-linearity between variables, we tested the correlation between scores on the SCS-R subscales. 
Finding only a weak Spearman's correlation coefficient of -0.3~\cite{AkogluCorrelation}, we proceeded with including both subscales as distinct dependent variables.

Additionally, each model included one context-relevant independent variable capturing attitudes towards similar situations involving \gii and NDII (RQ3b). For example, the model for \textsf{creation} included attitudes towards the creation of \gii within an intimate partnership as an independent variable and the model for \textsf{private\_sharing} included attitudes towards the indented recipient of \gii sharing it privately outside the relationship, without consent. During initial analysis, we decided to bucket these attitude items into ``unacceptable'' and ``not unacceptable'' to increase our statistical power. Lastly participant gender (bucketed into men and minoritized genders, see below) was included to address RQ4.

\aigncii is a form of image-based sexual abuse and tech-facilitated gender-based violence, which is predominantly, though not exclusively, perpetrated by cisgender men targeting cisgender women, transgender people, and/or non-binary people~\cite{DunnTFGBV, DunnDeepfakeVictims, McGlynnIBSA, wirtz2020transgender, McGlynnIBSAHarms}.
While research continues to investigate gendered proportions of perpetration and victimization---one report finds that most online \aigncii targeted women~\cite{AjderDeeptrace}, another report finds that men were more likely to report \aigncii victimization than women~\cite{UmbachAttitudes}---attitudes are nevertheless informed by the broader dynamics of gender-based violence.
Thus, mens' attitudes of \aigncii may differ from the attitudes of people who are not men.
In order to increase statistical power, we grouped people who were not men together, i.e., women, agender, or non-binary individuals and refer to this group as ``marginalized genders.''\footnote{In our survey, we did not ask whether respondents were transgender, so our sample of men includes transgender and cisgender men.}
Further, we only had 8 respondents who self-identified as agender, or non-binary; we bucketed them with women to include their responses in our quantitative analyses, rather than dropping the responses entirely. 
Additionally, we ran statistical models for `women' and `men', excluding participants outside this gender binary, which are similar and lead to the same conclusions (see Appendix~\ref{ap:additionalModels}).

To further examine the contextual factors' effect on acceptability (RQ2), another CLMM was built by adding interactions terms between \textsf{intent} and \textsf{action} as well as \textsf{intent} and \textsf{creator} to the original model for \textsf{creation}. To  examine the effect of participant gender on attitudes towards synthetic imagery (RQ4), five additional models were built by expanding the original models to include interactions terms between gender and each vignette factor. Of the expanded models, only the \textsf{creation} model showed statistically significant interaction effects ($p<0.05$) and thus was selected for further analysis. To compare acceptability across the actions of \textsf{creation}, \textsf{private\_sharing}, \textsf{public\_sharing}, \textsf{resharing}, and \textsf{seeking\_out}, another model was built with acceptability rating as the dependent variable and these actions as the independent variable.

\heading{Qualitative analysis.} We analyzed respondents' open-text rationale for their acceptability rating for the \textsf{creation} of the synthetic video using a coding reliability approach~\cite{BraunTA}.
The dataset was divided into two subsets, justifications for and against acceptability. Two researchers familiarized themselves with all rationales and generated an initial set of codes. The researchers compared and discussed codes to establish a final codebook (Appendix~\ref{ap:qualitative-codebook}). In line with qualitative research perspectives on the limitations of multiple coders~\cite{ArmstrongCoding, MorseCoding}, a single researcher performed the entire coding process for consistency and to preserve interpretive nuance~\cite{ElliotCoding}. A second researcher reviewed the codebook as well as 50 random responses from each subset in order to balance researcher subjectivity with thoroughness~\cite{WhittemoreCoding}. 

\subsection{Other considerations}

\headingg{Ethical considerations.}
\label{sec:ethics}
This study was deemed exempt by our IRB. However, ethical considerations extend beyond regulatory compliance~\cite{bhalerao2022ethical}. As vignettes describe non-consensual creation and sharing of intimate imagery, we were concerned about potential harm from placing respondents into hypothetical victimization scenarios, especially for those who have experienced image-based sexual abuse or sexual violence.

Consulting subject-area experts with training in clinical psychology and sexual trauma, we took the following steps for harm reduction: (1) surfacing in the consent form that the vignettes described synthetic media being created of the respondent,
(2) asking for re-consent after defining generative AI,
(3) including `prefer not to answer' option for all questions about intimate images, and 
(4) including contact information for IBSA support organizations at the end of the survey.
We also provided support resources for members of the research team who analyzed open-ended survey responses.

\heading{Positionality statement.}
Recognizing the inherent subjectivity in research, we acknowledge that our positionality as researchers shapes our approach to this work~\cite{HolmesPositionality, BardzellPositionality, BraunPositionality}. We bring varied perspectives informed by our distinct social, cultural, disciplinary, and ideological contexts. Our research team consists of three cisgender women and one cisgender man who are all researchers in security and privacy. As our team composition does not fully reflect the diversity of identities among our study respondents, there may be limitations in our thematic analysis and interpretation of the collected data.

\heading{Limitations.}
While surveys offer valuable insights, there are inherent limitations to using them. We prioritized reducing survey fatigue by pre-testing and piloting our survey. To minimize social desirability bias, we emphasized that each response about acceptability was based solely on the respondent's personal opinions. Our data is limited to the attitudes and justifications respondents were willing to report.

Crowdworking platforms offer access to large and diverse populations and are frequently used to elicit security and privacy attitudes~\cite{EmamiSmartCities, WeiGender, RedmilesGeneralize}; we chose Prolific for its higher data quality compared to other platforms~\cite{Prolific, PeerProlific}.
Anticipating that attitudes towards \aigncii vary by country, we chose to recruit solely in the US, which likely limits generalizability.

As noted in Section~\ref{sec:methodology:data-analysis:quant}, our survey instrument did not record transgender identities. As a result, our analysis may not fully capture the experiences of transgender individuals.

Additionally, as a formative study, we chose to explore specific factors (e.g., gender, contexts) rather than formulate uninformed hypotheses.
\section{Results}
\label{sec:results}
To quantitatively analyze the 315 survey responses, we built eight CLMMs (see Section~\ref{sec:methodology:data-analysis:quant}). The complete regression results for five, including the odds ratio (OR), confidence interval, and $p$-value range for each independent variable, are in Table~\ref{tab:FullRegression} 
\ifnum\confversion=1
(see the extended arXiv version for visualization~\cite{full}).
\else
(see Appendix~\ref{ap:or-graphs}, Figure~\ref{fig:ORGraphs} for visualization).
\fi
Where models with interactions are used (Table~\ref{tab:contextualInteractionRegressionTable} and Table~\ref{tab:genderInteractionRegressionTable}), only the models for \textsf{creation} had significant interaction terms and thus were selected for analysis.

Additionally, we conducted thematic analysis of the 861 open-response explanations of why participants found the \textsf{creation} of synthetic media in each vignette either acceptable or unacceptable. Aligned with qualitative methods, our analysis aimed to surface general themes about participants' attitudes, rather than quantify their prevalence. Accordingly, we report the appearance of themes using the following terminology: a few (less than 25\%), some (25-45\%), about half (45-55\%), most (55-75\%), and almost all (75-100\%). 
When providing participant quotes, we refer to each participant with the letter `P' followed by their unique participant number and specify the vignette they were responding to. 
\ifnum\confversion=1
Visualizations for the distributions of codes over vignettes and \textsf{action}s are available in the extended arXiv version of this paper~\cite{full}.
\else
Figure~\ref{fig:acceptable-code-distribution} and Figure~\ref{fig:unacceptable-code-distribution} in Appendices~\ref{ap:acceptable-code-heatmaps} visualize distributions of codes over vignettes and \textsf{action}s.
\fi
In some figures and this section, vignettes are referenced by their ID (e.g., V5) and the factor description \textsf{creator}/\textsf{action}/\textsf{intent} (see Table~\ref{tab:fullVignetteTable}).

In our results, we use \textit{synthetic} media to refer to media that is AI-generated, e.g., deepfakes, and \textit{\aigncii} to refer to synthetic media that are specifically intimate imagery.

\setlength{\textfloatsep}{8pt}
\begin{figure*}[t]
  \centering
    \includegraphics[width=1\textwidth]{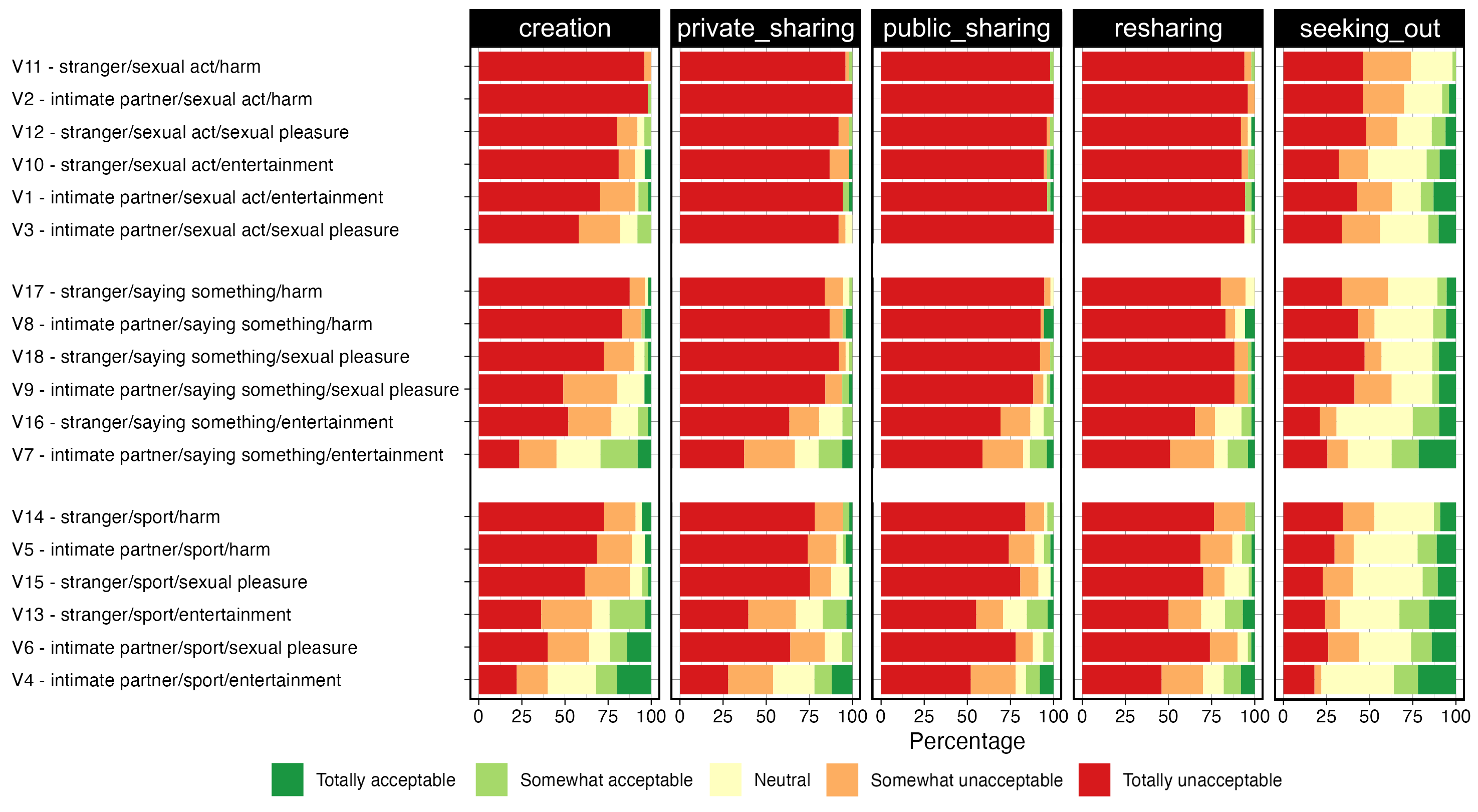}
  \caption{Respondents' perceptions of acceptability across all vignettes; each vignette is defined by the \textsf{creator} / \textsf{action} / \textsf{intent}. Vignettes are grouped by action and ordered (from bottom to top) by increasing unacceptability of creation.}
  \label{fig:AcceptabilityByAction}
\end{figure*}

\begin{table*}[t]
    \centering
    \footnotesize
    \renewcommand{\arraystretch}{2.25}
    \newcolumntype{Y}{>{\centering\arraybackslash}X}
    \newcolumntype{M}[1]{>{\centering\arraybackslash}m{#1}} 
    \begin{tabularx}{\textwidth}{M{0.3cm} M{6cm} *{5}{Y}}
    \toprule
        & & \textsf{creation} 
        & \textsf{private\_sharing}  
        & \textsf{public\_sharing}
        & \textsf{resharing}
        & \textsf{seeking\_out}  \\
    \midrule
    \multirow{4}{*}{\rotatebox[origin=c]{90}{\textbf{Intercepts\hspace{5pt}}}} 
        & Totally unacceptable | Somewhat unacceptable
        & \shortstack{5.13 \\ $[0.42,62.94]$ } 
        & \shortstack{1.19 \\ $[0.29, 115.33]$ } 
        & \shortstack{29.21 \\ $[0.86, 987.2]$ } 
        & \shortstack{1.02 \\ $[0.04, 24.7]$ } 
        & \cellcolor{yellow!30}\shortstack{0.03* \\ $[0, 0.8]$ } \\
        & Somewhat unacceptable | Neutral
        & \cellcolor{yellow!30}\shortstack{29.38** \\ $[2.35, 367.77]$ } 
        & \shortstack{5.82 \\ $[0.29, 115.33]$ } 
        & \cellcolor{yellow!30}\shortstack{122.87** \\ $[3.5, 4318.76]$ } 
        & \shortstack{4.72 \\ $[0.19, 114.86]$ } 
        & \shortstack{0.16 \\ $[0.01, 4.42]$ } \\
        & Neutral | Somewhat acceptable
        & \cellcolor{yellow!30}\shortstack{99.64*** \\ $[7.82, 1269.53]$ } 
        & \shortstack{18.57 \\ $[0.93, 370.82]$ } 
        & \cellcolor{yellow!30}\shortstack{289.97** \\ $[8.06, 10433.52]$ } 
        & \shortstack{15.28 \\ $[0.62, 374.82]$ } 
        & \shortstack{5.29 \\ $[0.19, 145.78]$ } \\
        & Somewhat acceptable | Totally acceptable 
        & \cellcolor{yellow!30}\shortstack{375.02*** \\ $[28.62, 4913.68]$ } 
        & \cellcolor{yellow!30}\shortstack{83.63** \\ $[4.08, 1713.2]$ } 
        & \cellcolor{yellow!30}\shortstack{1481.83*** \\ $[38.75, 56663.78]$ } 
        & \cellcolor{yellow!30}\shortstack{70.09* \\ $[2.78, 1767.14]$ } 
        & \shortstack{25.60 \\ $[0.92, 710.87]$ } \\
    \midrule
    \multirow{5}{*}{\rotatebox[origin=c]{90}{\textbf{Controlled IVs\hspace{5pt}}}} 
        & \textsf{creator} (Intimate partner) 
        & \cellcolor{yellow!30}\shortstack{3.24*** \\ $[2.23, 4.71]$ } 
        & \cellcolor{yellow!30}\shortstack{1.69* \\ $[1.13, 2.55]$ } 
        & \shortstack{1.47 \\ $[0.9, 2.4]$ } 
        & \shortstack{1.00 \\ $[0.65, 1.53]$ } 
        & \shortstack{1.11 \\ $[0.8, 1.53]$ } \\
        & \textsf{action} (Sport) 
        & \cellcolor{yellow!30}\shortstack{13.39*** \\ $[7.96, 22.52]$ } 
        & \cellcolor{yellow!30}\shortstack{34.72*** \\ $[16.76, 71.92]$ } 
        & \cellcolor{yellow!30}\shortstack{66.61*** \\ $[22.75, 19504]$ } 
        & \cellcolor{yellow!30}\shortstack{32.36*** \\ $[15.12, 69.25]$ } 
        & \cellcolor{yellow!30}\shortstack{7.26*** \\ $[4.73, 11.15]$ } \\
        & \textsf{action} (Saying something)  
        & \cellcolor{yellow!30}\shortstack{5.44*** \\ $[3.27, 9.05]$ } 
        & \cellcolor{yellow!30}\shortstack{11.01*** \\ $[5.45, 22.23]$ } 
        & \cellcolor{yellow!30}\shortstack{19.49*** \\ $[6.91, 54.94]$ } 
        & \cellcolor{yellow!30}\shortstack{12.47*** \\ $[5.92, 26.29]$ } 
        & \cellcolor{yellow!30}\shortstack{3.40*** \\ $[2.21, 5.22]$ } \\
        & \textsf{intent} (Entertainment) 
        & \cellcolor{yellow!30}\shortstack{18.92*** \\ $[11.03, 32.46]$ } 
        & \cellcolor{yellow!30}\shortstack{11.49*** \\ $[6.59, 20.05]$ } 
        & \cellcolor{yellow!30}\shortstack{10.57*** \\ $[5.39, 20.73]$ } 
        & \cellcolor{yellow!30}\shortstack{5.51*** \\ $[3.18, 9.56]$ } 
        & \cellcolor{yellow!30}\shortstack{4.94*** \\ $[3.25, 7.49]$ } \\
        & \textsf{intent} (Sexual pleasure)
        & \cellcolor{yellow!30}\shortstack{7.42*** \\ $[4.42, 12.47]$ } 
        & \shortstack{1.35 \\ $[0.77, 2.37]$ } 
        & \shortstack{1.15 \\ $[0.58, 2.28]$ } 
        & \shortstack{0.92 \\ $[0.52, 1.63]$ } 
        & \shortstack{1.37 \\ $[0.92, 2.04]$ } \\
    \midrule
    \multirow{4}{*}{\rotatebox[origin=c]{90}{\textbf{Uncontrolled IVs\hspace{5pt}}}} 
        & Gender (Man) 
        & \cellcolor{yellow!30}\shortstack{2.45*** \\ $[1.45, 4.15]$ } 
        & \cellcolor{yellow!30}\shortstack{2.12** \\ $[1.21, 3.7]$ } 
        & \shortstack{1.77 \\ $[0.88, 3.55]$ } 
        & \shortstack{1.41 \\ $[0.75, 2.66]$ }
        & \shortstack{1.51 \\ $[0.76, 2.99]$ } \\
        & \gii \& NDII attitudes (Unacceptable) 
        & \cellcolor{yellow!30}\shortstack{0.21* \\ $[0.05, 0.84]$ } 
        & \cellcolor{yellow!30}\shortstack{0.08* \\ $[0.01, 0.4]$ } 
        & \cellcolor{yellow!30}\shortstack{0.09** \\ $[0.02, 0.41]$ } 
        & \cellcolor{yellow!30}\shortstack{0.01*** \\ $[0, 0.05]$ } 
        & \cellcolor{yellow!30}\shortstack{0.01*** \\ $[0.01, 0.03]$ } \\
        & SCS-R2
        & \cellcolor{yellow!30}\shortstack{0.53*** \\ $[0.39, 0.72]$ } 
        & \cellcolor{yellow!30}\shortstack{0.55*** \\ $[0.4, 0.77]$ } 
        & \cellcolor{yellow!30}\shortstack{0.64* \\ $[0.42, 0.96]$ } 
        & \shortstack{0.76 \\ $[0.52, 1.1]$ }
        & \shortstack{0.73 \\ $[0.48, 1.1]$ } \\
        & SCS-R4
        & \shortstack{1.06 \\ $[0.82, 1.36]$ } 
        & \shortstack{1.10 \\ $[0.84, 1.44]$ } 
        & \shortstack{1.30 \\ $[0.92, 1.82]$ } 
        & \shortstack{1.27 \\ $[0.93, 1.72]$ }
        & \shortstack{1.14 \\ $[0.82, 1.59]$ } \\
    \bottomrule
    \end{tabularx}
    \caption{\small{Results from regressions exploring the relationship between scenario acceptability (first row, intercepts), contextual factors (second row, controlled IVs), and personal factors (third row, uncontrolled IVs). Each column represents the output of one regression model. Numeric cells list the odds ratio (OR) and the 95\% confidence interval.  Reference levels: \textsf{creator} (stranger), \textsf{action} (sexual act), \textsf{intent} (harm), gender (marginalized genders), \gii \& NDII attitudes (acceptable). Significance of OR: $p<0.05$ = \colorbox{yellow!30}{*}, $p<0.01$ = \colorbox{yellow!30}{**}, and $p<0.001$ = \colorbox{yellow!30}{***}.}} 
    \label{tab:FullRegression}
\end{table*}

\subsection{General Attitudes (RQ1)}
\label{sec:results:general}

People generally found the \textsf{creation} of synthetic media unacceptable, with a median percentage of somewhat or totally unacceptable ratings across all scenarios of 89.54\%. They perceived any sharing of these media as even more unacceptable: 94.39\% for \textsf{private\_sharing}, 94.44\% for \textsf{public\_sharing}, 94.22\% for \textsf{resharing}. Attitudes were more mixed regarding \textsf{seeking\_out} such media, however (52.78\%). 
The results of the regression examining the acceptability rating as the dependent variable with these actions as the independent variable, support these results statistically (see Table~\ref{tab:metaactionRegression} in Appendix~\ref{ap:metaaction-regression} for full results): Across scenarios and controlling for within-subject variation we observe that \textsf{private\_sharing} ($\text{OR}=0.47, p<0.001$), \textsf{public\_sharing} ($\text{OR}=0.26, p<0.001$), and \textsf{resharing} ($\text{OR}=0.42, p<0.001$) are significantly less acceptable than \textsf{creation} (the reference level). \textsf{seeking\_out} ($\text{OR}=5.43, p<0.001$) is significantly more acceptable than \textsf{creation}.

Figure \ref{fig:AcceptabilityByAction} illustrates these results visually, depicting perceived acceptability across \textsf{creation}, \textsf{private\_sharing}, \textsf{public\_sharing}, \textsf{resharing}, and \textsf{seeking\_out} for all vignettes. The rightmost column (\textsf{seeking\_out}) exhibits far more variance in attitudes than the columns to the left, although these variances differ depending on the depicted action, as we investigate next.

\heading{\aigncii perceived as less acceptable than other synthetic media not depicting sexual acts.}
\label{sec:results:general:AIGNCII-less-acceptable}
While people broadly found \textsf{creation} and any form of \textsf{sharing} of synthetic media unacceptable, this was particularly true for \aigncii (RQ2). 
Across \textsf{creation}, \textsf{private\_sharing}, \textsf{public\_sharing}, and \textsf{resharing} contexts, scenarios in which the \textsf{action} was playing a sport or saying something, as opposed to performing a sexual act, were rated as more acceptable by participants ($\text{OR}>7, p<0.001$ for all models in Table~\ref{tab:FullRegression}).

Turning again to Figure~\ref{fig:AcceptabilityByAction}, we observe this effect clearly. 
Regarding \textsf{creation}, the least unacceptable scenario depicting a sexual act was V3 -- an intimate partner non-consensually creating synthetic media of the participant engaged in a sexual act for their sexual pleasure -- 82\% of respondents found this scenario to be somewhat or totally unacceptable.
\footnote{The potential for flattery within a relationship (see Section~\ref{sec:results:contextual:fantasy}) may explain why this was lower than the median across all vignettes.}
The most accepted scenario depicting the participant saying something (V7) -- an intimate partner non-consensually creating synthetic media of the participant saying something for entertainment -- was considered unacceptable by about half of participants (45.1\%). The most acceptable scenario in our entire survey (V4), which depicted an intimate partner non-consensually creating synthetic media of the participant playing a sport was considered unacceptable by just a third (32\%) of participants.

\begin{figure}[t]
  \centering
  \includegraphics[width=0.45\textwidth]{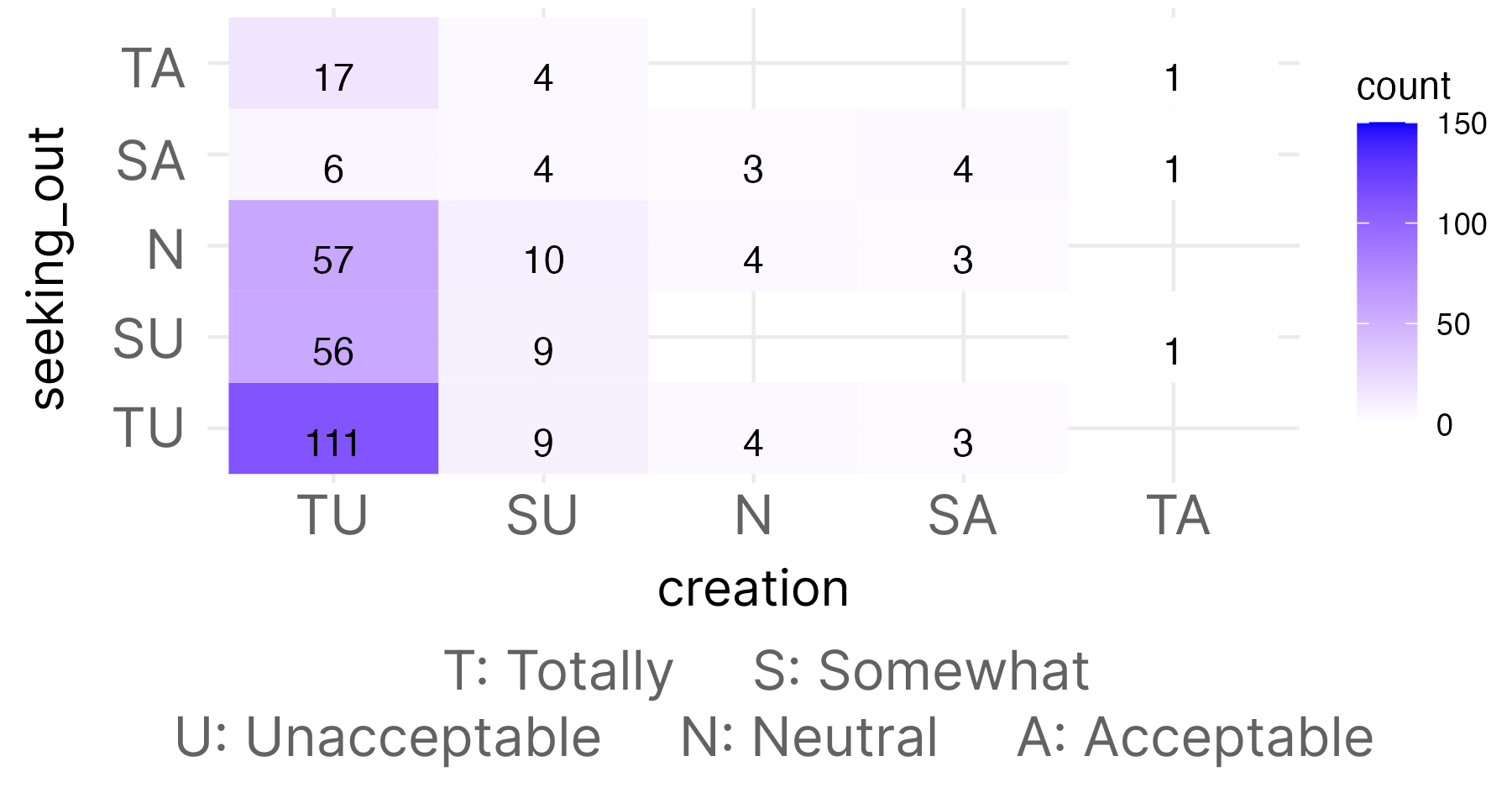}
  \caption{Heatmap of acceptability for \textsf{creation} and \textsf{seeking\_out} when the 
  \textsf{action} is performing a sexual act.}
  \label{fig:creationVsSeekingOut}
\end{figure}

\textsf{seeking\_out} \aigncii was also viewed as less acceptable than \textsf{seeking\_out} other forms of synthetic content ($\text{OR}>3, p<0.001$; Table~\ref{tab:FullRegression}). However, when comparing \textsf{seeking\_out} \aigncii to creating it, it is still more acceptable than \textsf{creation} as illustrated by Figure~\ref{fig:creationVsSeekingOut}.

\heading{Portrayed action relates to perceived harm.}
\label{sec:results:general:harms}
When explaining their perception of a scenario, some participants remarked on potential harm to their reputation or lack thereof to explain why they viewed \textsf{creation} as acceptable or unacceptable. Lack of harm was the most common reason for finding synthetic media \textsf{creation} acceptable, typically when that media depicted the subject playing a sport. For example:
\begin{quote}
    \itshape
    There is nothing sexual\ldots that i woul[dn']t want the public to know/see (P50, V13 - stranger/sport/entertainment).
    \end{quote}
On the other hand, when discussing AIG-NCII or depictions of them saying something they did not, some participants remarked on the potential harms of that content:
\begin{quote}
    \itshape
        Sexual act will tarnish my image in the society (P193, V10 - stranger/sexual act/entertainment).
        
        AI can seem realistic. Whatever they have me saying could be used against me in a variety of situations (P32, V16 - stranger/saying something/entertainment).
    \end{quote}
    
Further, when the \textsf{action} was performing a sexual act, a few participants also observed that the \textsf{creation} of \aigncii wrong because --- even if synthetic --- the images violated the sanctity of their bodies, e.g.: 
\begin{quote}
    \itshape
    It’s a violation of my body and it is disrespectful (P49, V10 - stranger/sexual act/entertainment). 

    I feel it's unacceptable to manipulate my image in such a way - my body and how it looks belongs to me (P195, V1 - intimate partner/sexual act/entertainment).
\end{quote}

Finally, while we only asked respondents to explain their judgements of (un)acceptability relating to media \textsf{creation} (Section~\ref{sec:methodology:survey-structure:vignette}), some mentioned the stage of media production (e.g., \textsf{creation} vs. any form of \textsf{sharing}) influenced the likelihood of harm and thus their perception of acceptability:
\begin{quote}
    \itshape
    It’s not harming me or blackmailing me or anything. As long as it doesn't get shared I think it's ok (P163, V3 - intimate partner/sexual act/sexual pleasure).
\end{quote}

\headingg{Some respondents call on morality, legality, and privacy to explain the unacceptability of synthetic media.}
A few participants justified the \textsf{creation} of synthetic media depicting them as unacceptable because it was amoral or unethical to create fake content without the subject's consent, e.g.,
\begin{quote}
    \itshape
    This is a false representation of me and highly unethical (P204, V16 - stranger/saying something/entertainment).

    I don’t think it is right to use a person[']s identity to say things that they didn’t say (P302, V16 - stranger/saying something/entertainment).
\end{quote}

While not specifically speaking to amorality, a few expressed sentiments of disgust often associated in psychological literature with intuitive responses to moral violations~\cite{giner2018makes}: that the \textsf{creation} of the content was `gross' (P50), `creepy' (P24), `weird' (P74), or `nasty' (P112). Such feelings were especially prevalent when the content was created by a stranger or the action depicted was incongruous with the intent (e.g., a stranger creating a video of someone playing a sport for sexual pleasure). We explore these variations based on contextual factors further in Section~\ref{sec:results:contextual:incongruent}. 

In a few other cases, participants referred to the \textsf{creation} of the media as illegal or compared it to a crime, despite the fact that no federal legal protections currently exist on \aigncii~\cite{williams2023legal}. Across all \textsf{action}s, a few participants called the act of \textsf{creation} slanderous, like P268 in response to V14 (stranger/sport/harm):
\begin{quote}
    \itshape
    They are using faked info to harm me. This is slander. 
\end{quote}
When the \textsf{action} was saying something, the \textsf{creation} was often compared to libel or fraud, e.g.,
\begin{quote}
    \itshape
    It seems like the equivalent of slander and fraud. If this were done in election ads, it would be disallowed/illegal (P253, V17 - stranger/saying something/harm).

    [I]t is never acceptable to lie. I would sue for libel (P259, V7 - intimate partner/saying something/entertainment).
\end{quote}
Specific to \aigncii, participants mentioned crimes of sexual violence,
\begin{quote}
    \itshape
    This scenario is harmful and akin to some form of sexual ha[r]assment or assault, especially done without knowledge (P212, V2 - intimate partner/sexual act/harm).
\end{quote}

Finally, a few respondents called the \textsf{creation} of synthetic media of them a privacy violation, e.g.:
\begin{quote}
    \itshape
    This completely violates my sense of privacy (P10, V2 - intimate partner/sexual act/harm).
    
    Creating an image of a person without their knowledge is a violation of privacy (P170, V6 - intimate partner/sport/sexual pleasure). 
\end{quote}
This attitude appeared relatively evenly and similarly in rationales across all \textsf{action}s.

\subsection{Role of contextual factors (RQ2)} 
\label{sec:results:contextual}
\begin{table}[t]
\renewcommand{\arraystretch}{1.2}
\centering
\footnotesize
\begin{tabular}{m{0.3cm}m{4.1cm}m{3cm}}
\toprule
& & OR; Confidence Interval \\
\midrule
\multirow{4}[0]{*}{\rotatebox[origin=c]{90}{\textbf{Intercepts\hspace{5pt}}}}
& \vspace{2pt} \shortstack[l]{Totally unacceptable | \\\;\;Somewhat unacceptable} & 7.51; [0.42, 134.86] \\
& \cellcolor{yellow!30}  Somewhat unacceptable | Neutral & \cellcolor{yellow!30}47.58; [2.61, 867.1]** \\
& \cellcolor{yellow!30} Neutral | Somewhat acceptable & \cellcolor{yellow!30}171.35; [9.26, 3169.4]*** \\
& \vspace{2pt} \cellcolor{yellow!30}\shortstack[l]{Somewhat acceptable | \\\;\;Totally acceptable} & \cellcolor{yellow!30}665.83; [35.15, 12613]*** \\
\midrule
\multirow{4}[0]{*}{\rotatebox[origin=c]{90}{\textbf{Controlled IVs\hspace{3pt}}}}
& \textsf{creator} (Intimate partner) & 1.38; [0.62, 3.04] \\
& \cellcolor{yellow!30} \textsf{action} (Sport) & \cellcolor{yellow!30}48.94; [11.43, 209.59]*** \\
& \cellcolor{yellow!30} \textsf{action} (Saying something) & \cellcolor{yellow!30}9.9; [2.24, 43.77]** \\
& \cellcolor{yellow!30} \textsf{intent} (Entertainment) & \cellcolor{yellow!30}13.14; [2.86, 60.3]*** \\
& \cellcolor{yellow!30} \textsf{intent} (Sexual pleasure) & \cellcolor{yellow!30}20.82; [4.53, 95.72]*** \\
\midrule
\multirow{4}[0]{*}{\rotatebox[origin=c]{90}{\textbf{\shortstack{Uncontrolled\\IVs}}}}
& \cellcolor{yellow!30} Gender (man) & \cellcolor{yellow!30} 2.64; [1.53, 4.57]*** \\
& \cellcolor{yellow!30} \gii \& NDII attitudes (Unacceptable) & \cellcolor{yellow!30}0.19; [0.04, 0.8]* \\
& \cellcolor{yellow!30} SCS-R2 & \cellcolor{yellow!30}0.51; [0.37, 0.71]*** \\
& SCS-R4 & 1.08; [0.83, 1.4] \\
\midrule
\multirow{4}[0]{*}{\rotatebox[origin=c]{90}{\textbf{Interaction Terms\hspace{24pt}}}}
& \cellcolor{yellow!30} \shortstack[l]{ \textsf{creator} (Intimate partner) \& \\\;\textsf{intent} (Entertainment)} &\cellcolor{yellow!30} 2.83; [1.07, 7.5]* \\
& \vspace{2pt} \cellcolor{yellow!30} \shortstack[l]{\textsf{creator} (Intimate partner) \& \\\;\textsf{intent} (Sexual pleasure)} &\cellcolor{yellow!30} 3.76; [1.38, 10.2]** \\
& \vspace{2pt} \shortstack[l]{\textsf{action} (Sport) \& \\\;\textsf{intent} (Entertainment)} & 0.72; [0.15, 3.56] \\
& \vspace{2pt} \shortstack[l]{\textsf{action} (Saying something) \& \\\; \textsf{intent} (Entertainment)} & 1.68; [0.33, 8.66] \\
& \vspace{2pt} \cellcolor{yellow!30} \shortstack[l]{\textsf{action} (Sport) \& \\\;\textsf{intent} (Sexual pleasure)}&
\cellcolor{yellow!30}0.08; [0.02, 0.4]** \\
& \vspace{2pt} \shortstack[l]{\textsf{action} (Saying something) \& \\\; \textsf{intent} (Sexual pleasure)} & 0.19; [0.04, 1.01] \\
\bottomrule
\end{tabular}
\caption{Results from a single regression exploring the relationship between the acceptability of \textsf{creation} (first row, intercepts), contextual factors (second row, controlled IVs), personal factors (third row, uncontrolled IVs), and interactions between \textsf{intent} and \textsf{creator} or \textsf{action} (fourth row, interaction terms). Reference levels: \textsf{creator} (stranger), \textsf{action} (sexual act), \textsf{intent} (harm), gender (marginalized genders), \gii \& NDII attitudes (acceptable). Significance of OR: $p<0.05$ = \colorbox{yellow!30}{*}, $p<0.01$ = \colorbox{yellow!30}{**}, and $p<0.001$ = \colorbox{yellow!30}{***}.}
\label{tab:contextualInteractionRegressionTable}
\end{table}
Consistent with the theory of contextual integrity~\cite{NissenbaumPrivacy}, we found that contextual factors strongly influenced both respondents' ratings of acceptability and their rationales. 
\vspace{3pt}

\headingg{It is more acceptable for intimate partners to create synthetic media than strangers, but only if they do not intend harm.} \label{sec:results:contextual:intimate-partner-no-harm} We observe from Table~\ref{tab:FullRegression} that across all scenarios, when the content \textsf{creator} was an intimate partner as opposed to a stranger, participants were more likely to find the \textsf{creation} ($\text{OR}=3.24, p<0.001$; Table~\ref{tab:FullRegression}) as well as the \textsf{private\_sharing} ($\text{OR}=1.69, p=0.01$; Table~\ref{tab:FullRegression}) of the synthetic imagery more acceptable (RQ2). 
However, when we consider interactions with the \textsf{intent} of the synthetic media (Table~\ref{tab:contextualInteractionRegressionTable}), we observe that there is no longer a significant relationship between \textsf{creator} and acceptability of \textsf{creation} and that there are three significant interactions between: (1) \textsf{creator} being an intimate partner and \textsf{intent} being entertainment ($\text{OR}=2.83, p=0.036$; Table~\ref{tab:contextualInteractionRegressionTable}), (2) \textsf{creator} being an intimate partner and \textsf{intent} being sexual pleasure ($\text{OR}=3.76, p=0.009$; Table~\ref{tab:contextualInteractionRegressionTable}), as well as between (3) \textsf{action} being playing a sport and \textsf{intent} being sexual pleasure ($\text{OR}=0.08, p=0.002$; Table~\ref{tab:contextualInteractionRegressionTable}), which we address later in this section.
Thus, our interaction model demonstrates a more nuanced answer to RQ2.
The main effect we observed in our original modeling for \textsf{creation} (without interactions) -- that intimate partners creating synthetic media is more acceptable -- was driven by attitudes that intimate partners creating synthetic media for non-harmful purposes is more acceptable. That is, if the \textsf{creator} is an intimate partner and the \textsf{intent} is entertainment ($\text{OR}=2.83, p=0.036$; Table~\ref{tab:contextualInteractionRegressionTable}) or sexual pleasure ($\text{OR}=3.76, p=0.009$; Table~\ref{tab:contextualInteractionRegressionTable}) the media \textsf{creation} is more acceptable. However, intimate partners creating media for the \textsf{intent} to harm is no more acceptable than a stranger doing so.

\ifnum\confversion=1
\headingg{Intimate partner trust related to explanations of (un)acceptability.}
\else
\heading{Intimate partner trust related to explanations of (un)acceptability.}
\fi
Some explanations for acceptability, like P211's response to V1 (intimate partner/sexual act/entertainment), reflected trust in a partner enabling acceptable \textsf{creation}:
\begin{quote}
    \itshape
    I feel if we are intimate, we're already engaging in similar acts. It's all in good sexual fun, as long as they don't distribute it or show anyone else.
\end{quote}
This exhibits a belief that an intimate relationship permits intimate media \textsf{creation} within it, whereas no such trust exists in relationships with strangers, increasing feelings of violation:
\begin{quote}
    \itshape
    The idea of somebody I don't know generating porn of me is insanely creepy (P24, V12 - stranger/sexual act/sexual pleasure)
\end{quote}

On the other hand, some explanations for unacceptability stated that the \textsf{creation} \textit{violated} intimate partner trust rather than being acceptable because of it, e.g.,
\begin{quote}
    \itshape
    I think this is just as worse because there is supposed to be a trust between people who are intimate and they completely broke that trust (P142, V3 - intimate partner/sexual act/sexual pleasure).
\end{quote}
About half of the rationales exhibiting this attitude were in response to the \textsf{creation} of synthetic media of sexual acts.

\heading{A few were flattered by the \textsf{creation} of material for sexual fantasy within an intimate partnership.} \label{sec:results:contextual:fantasy} In scenarios where synthetically generated media was created for sexual gratification by an intimate partner, a few participants reported feelings of being flattered by its production, e.g.,
\begin{quote}
    \itshape
    The content she generated sounds cool and indicates she’s attracted to me (P65, V6 - intimate partner/sport/sexual pleasure).

    I don’t care what my intimate partners choose to do. I would be flattered (P65, V9 - intimate partner/saying something/sexual pleasure).
\end{quote}

A few noted that they couldn't control the sexual fantasies of others, regardless of whether they were in a relationship: 
\begin{quote}
    \itshape
 I don't particularly like that and I would prefer they don't do it, but I can't stop them from fantasizing about me in their own head.  I can't stop them from writing down their fantasies on paper or drawing a picture (P188, V12 - stranger/sexual act/sexual pleasure).
\end{quote}
While others expressed that, in the context of an intimate relationship, they would prefer to engage in their partner's fantasy instead:
\begin{quote}
    \itshape
    It's a bit bizarre and strange. I'd rather I actually perform this act instead of a fake AI version of me doing so (P165, V1 - intimate partner/sexual act/entertainment).
\end{quote}

\noindent\textbf{Intent impacts acceptability ratings differently depending on stage in the media pipeline.} We observe from Table~\ref{tab:FullRegression} that regardless of the \textsf{creator} of the media, respondents rated as more acceptable those scenarios where synthetic videos were created, shared, and sought out for entertainment vs.\ with intent to harm ($\text{OR}>4, p<0.001$; Table~\ref{tab:FullRegression}). Respondents also found \textsf{creation} of synthetic videos with the intent of bringing the creator sexual pleasure more acceptable than \textsf{creation} with the intent to harm the subject. However, respondents did not rate the acceptability of any form of \textsf{sharing} or \textsf{seeking\_out} synthetic videos created with the intent of sexual pleasure differently from the acceptability of \textsf{sharing} or \textsf{seeking\_out} synthetic videos created with the intent to harm.

\heading{Incongruent actions and intentions increase unacceptability.} \label{sec:results:contextual:incongruent} Considering our interaction model, we find that these results hold but observe a further effect: incongruence between the \textsf{action} and the \textsf{intent} -- even for \textsf{action}s and \textsf{intent}s viewed as generally more acceptable -- reduce attitudes of acceptability. For example, while creating media depicting the subject playing a sport was overall more acceptable than depicting them engaged in a sexual act and depictions of any action for sexual pleasure were more acceptable than depictions for harm, depicting someone playing a sport with the intent of sexual pleasure was less acceptable than depicting a more congruous action (saying something, a sexual act) with the same intent. A few participants shared explanations for the (un)acceptability of synthetic media \textsf{creation} that support this finding, for example:
\begin{quote}
    \itshape
    That’s really creepy! It just grosses me out, even if it’s just sports. (P25, V15 - stranger/sport/sexual pleasure)
\end{quote}

\subsection{Role of sexual consent \& content attitudes (RQ3)}
\label{sec:results:sexual-consent-content}

\headingg{Attitudes toward establishing sexual consent offline relate to attitudes toward AI media generation and sharing.} \label{sec:results:sexual-consent-content:SCSR}
We used the second subscale from the SCS-R to measure attitudes towards establishing sexual consent~\cite{HumphreysSCSR} and answer RQ3a.
Those who scored higher on SCS-R2, indicating more positive attitudes toward establishing sexual consent, were less likely to rate non-consensual \textsf{creation}, \textsf{private\_sharing} or \textsf{public\_sharing} of synthetic content as acceptable ($\text{OR}<0.7, p<0.005$ for these models; in Table~\ref{tab:FullRegression}).

\heading{The most common explanation for finding synthetic media \textsf{creation} unacceptable is lack of consent.} For example, P19 remarked in response to V3 ( intimate partner/sexual act/sexual pleasure) that:
\begin{quote}
     \itshape
     No content should be made in someone else's likeness without their consent.
\end{quote}
The fourth SCS-R subscale measures attitudes towards consent norms specifically in the context of relationships and sexual activity~\cite{HumphreysSCSR}. Scores on this subscale did not significantly affect any models.

\begin{figure}
   \centering
   \includegraphics[width=0.5\textwidth]{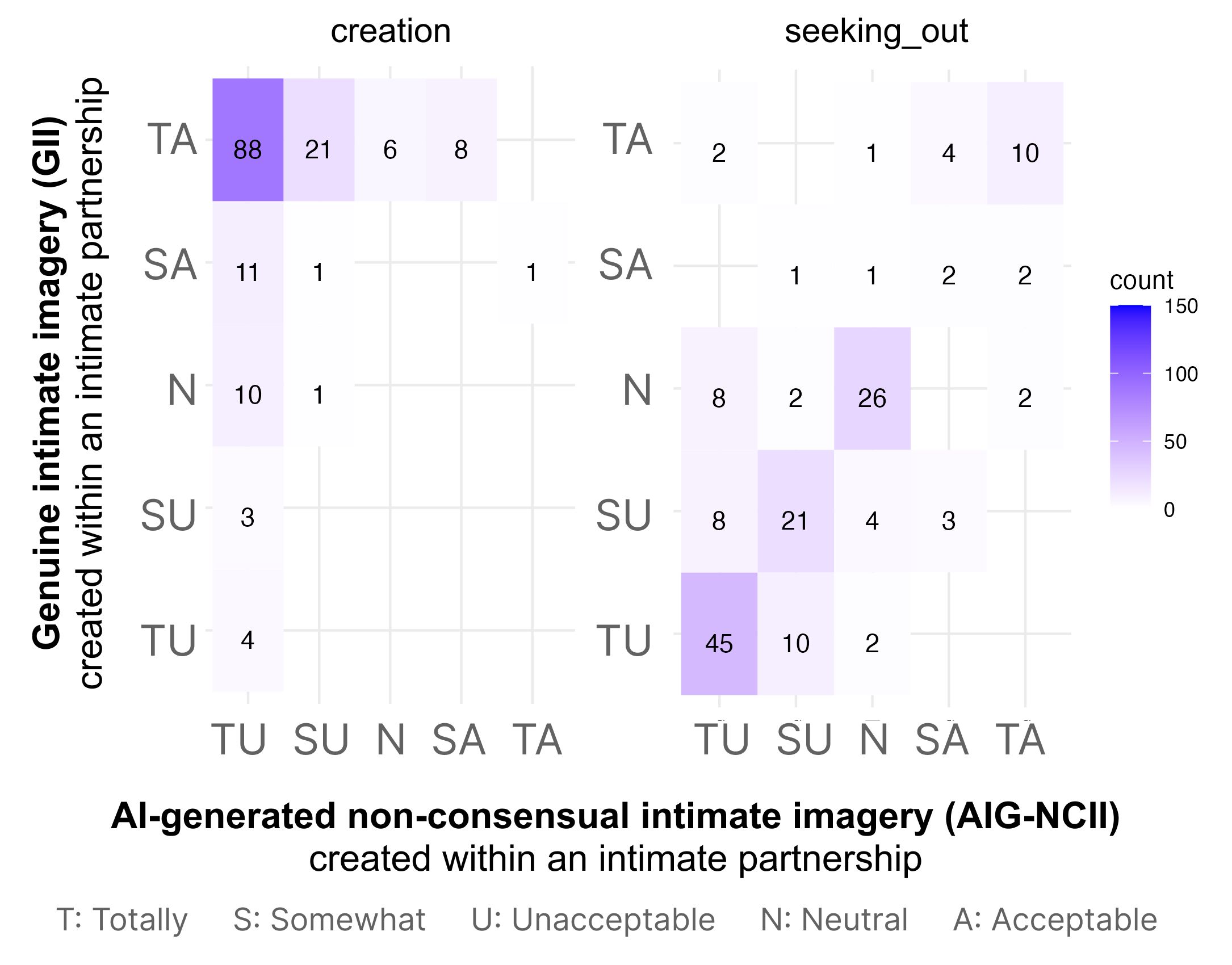}
   \caption{Heatmaps comparing acceptability of \textsf{creation} and \textsf{seeking\_out} for AIG-NCII to similar actions for \gii also created in an intimate relationship. See 
   \ifnum\confversion=1
    the extended arXiv version~\cite{full}
    \else
    Appendix~\ref{ap:GIIHeatmaps-full}, Figure~\ref{fig:GIIHeatmaps-full}
    \fi 
   for heatmaps including all forms of \textsf{sharing}.}
   \label{fig:AIIHeatmap}
\end{figure}

\ifnum\confversion=1
\headingg{Attitudes toward consensually-created genuine intimate imagery as well as NDII correlate with acceptance of synthetic videos including AIG-NCII.}
\else
\heading{Attitudes toward consensually-created genuine intimate imagery as well as NDII correlate with acceptance of synthetic videos including AIG-NCII.}
\fi
In addressing RQ3b, we sought to understand whether and how attitudes toward genuine, consensually-created intimate imagery related to attitudes toward synthetic, non-consensually created media. 

Those who found consensual creation of genuine intimate imagery (\gii) in an intimate relationship (somewhat or completely) unacceptable were also less likely to find non-consensual, synthetic \textsf{creation} of media depicting them acceptable, regardless of the act depicted ($\text{OR}=0.21, p=0.028$; Table~\ref{tab:FullRegression}). Those who found further sharing of GII without the original sender's consent -- i.e., non-consensual distribution of intimate imagery or NDII -- unacceptable were also less likely to find sharing of synthetic videos depicting them acceptable ($\text{OR}<0.1, p<0.05$ for \textsf{private\_sharing}, \textsf{public\_sharing}, and \textsf{resharing}; Table~\ref{tab:FullRegression}). Finally, those who considered seeking out NDII unacceptable were less likely to find \textsf{seeking\_out} synthetic videos acceptable ($\text{OR}=0.01, p<0.001$; Table~\ref{tab:FullRegression}).

In Figure~\ref{fig:AIIHeatmap}, we observe that over three fourths of participants who responded to a vignette involving \aigncii in the context of an intimate relations found consensual \gii creation within an intimate partnership totally acceptable (116/154), while none viewed non-consensual synthetic intimate media \textsf{creation} within an intimate partnership as totally acceptable. A key difference is that the \gii creation scenario implies awareness and consent, while the synthetic media vignettes explicitly do not. 
Considering non-consensual sharing, a majority of respondents viewed \textsf{private\_sharing} (140/153\footnote{Denominators vary because some participants preferred not to answer certain questions about synthetic and/or genuine intimate imagery.}), \textsf{public\_sharing} (145/151), and \textsf{resharing} (139/153) as totally unacceptable for both media types.
There was less consensus on \textsf{seeking\_out} non-consensually publicized synthetic and non-synthetic imagery, with only some (45/154) finding it totally unacceptable for both.

\ifnum\confversion=1
\vspace{-0.6em}
\else

\fi 
\subsection{Role of gender (RQ4)} \label{sec:results:gender}
\begin{table}[t]
\renewcommand{\arraystretch}{1.2}
\centering
\footnotesize
\begin{tabular}{m{0.3cm}m{4.1cm}m{3cm}}
\toprule
& & OR; Confidence Interval \\
\midrule
\multirow{4}[0]{*}{\rotatebox[origin=c]{90}{\textbf{Intercepts\hspace{5pt}}}}
& \vspace{2pt} \shortstack[l]{Totally unacceptable | \\\;\;Somewhat unacceptable} & 4.77; [0.33, 69.72] \\
& \cellcolor{yellow!30}  Somewhat unacceptable | Neutral & \cellcolor{yellow!30}29.29; \cellcolor{yellow!30}[1.97, 435.77]* \\
& \cellcolor{yellow!30} Neutral | Somewhat acceptable & \cellcolor{yellow!30}104.04; \cellcolor{yellow!30}[6.89, 1570.59]*** \\
& \vspace{2pt} \cellcolor{yellow!30}\shortstack[l]{Somewhat acceptable | \\\;\;Totally acceptable} & \cellcolor{yellow!30}410.88; [26.53, 6363.7]*** \\
\midrule
\multirow{4}[0]{*}{\rotatebox[origin=c]{90}{\textbf{Controlled IVs\hspace{3pt}}}}
& \cellcolor{yellow!30} \textsf{creator} (Intimate partner) & \cellcolor{yellow!30}1.75; [1.04, 2.96]* \\
& \cellcolor{yellow!30} \textsf{action} (Sport) & \cellcolor{yellow!30}17.58;[8.06, 38.33]*** \\
& \cellcolor{yellow!30} \textsf{action} (Saying something) & \cellcolor{yellow!30}10.68; [4.82, 23.65]*** \\
& \cellcolor{yellow!30} \textsf{intent} (Entertainment) & \cellcolor{yellow!30}20.05; [9.39, 42.85]*** \\
& \cellcolor{yellow!30} \textsf{intent} (Sexual pleasure) & \cellcolor{yellow!30}4.90; [2.32, 10.33]*** \\
\midrule
\multirow{4}[0]{*}{\rotatebox[origin=c]{90}{\textbf{\shortstack{Uncontrolled\\IVs}}}}
& Gender (man) & 1.54; [0.43, 5.61] \\
& \cellcolor{yellow!30} \gii \& NDII attitudes (Unacceptable) & \cellcolor{yellow!30}0.2; [0.05, 0.84]* \\
& \cellcolor{yellow!30} SCS-R2 & \cellcolor{yellow!30}0.52; [0.38, 0.72]*** \\
& SCS-R4 & 1.08; [0.83, 1.41] \\
\midrule
\multirow{4}[0]{*}{\rotatebox[origin=c]{90}{\textbf{Interaction Terms\hspace{24pt}}}}
& \shortstack[l]{\textsf{action} (Sport) \& \\\;Gender (Man)} & 0.77; [0.29,2.02] \\
& \vspace{2pt} \cellcolor{yellow!30} \shortstack[l]{\textsf{action} (Saying something) \& \\\;Gender (Man)} & \cellcolor{yellow!30} 0.32; [0.12, 0.88]* \\
& \vspace{2pt} \shortstack[l]{\textsf{intent} (Entertainment) \& \\\;Gender (Man)} & 1; [0.38, 2.61] \\
& \vspace{2pt} \shortstack[l]{\textsf{intent} (Sexual pleasure) \& \\\;Gender (Man)} & 2.36; [0.87, 6.43] \\
& \vspace{2pt} \cellcolor{yellow!30} \shortstack[l]{\textsf{creator} (Intimate partner) \& \\\;Gender (Man)}& \cellcolor{yellow!30}3.59; [1.71, 7.5]*** \\
\bottomrule
\end{tabular}
\caption{Results from a single regression exploring the relationship between scenario acceptability for \textsf{creation} (first row, intercepts), contextual factors (second row, controlled IVs), personal factors (third row, uncontrolled IVs), and interactions between gender and contextual factors (third row, interaction terms). Reference levels: \textsf{creator} (stranger), \textsf{action} (sexual act), \textsf{intent} (harm), gender (marginalized genders), \gii \& NDII attitudes (acceptable). Significance of OR: $p<0.05$ = \colorbox{yellow!30}{*}, $p<0.01$ = \colorbox{yellow!30}{**}, and $p<0.001$ = \colorbox{yellow!30}{***}.}
\label{tab:genderInteractionRegressionTable}
\end{table}
For quantitative analysis, we binned respondents by gender into men and marginalized genders (see Section~\ref{sec:methodology:data-analysis:quant}). Across scenarios, men were more likely to rate the \textsf{creation} ($\text{OR}=2.45, p<0.001$; Table~\ref{tab:FullRegression}) and \textsf{private\_sharing} ($\text{OR}=2.12, p=0.009$; Table~\ref{tab:FullRegression}) more acceptable than people with a marginalized gender.

\heading{Men view synthetic media depicting them engaged in a sexual act more acceptable than others.} To further examine the role of gender identity in shaping attitudes towards non-consensual synthetic imagery \textsf{creation}, we performed an additional regression that included interaction terms between participant gender and each vignette factor (Table~\ref{tab:genderInteractionRegressionTable}).
We observe that the main effect of gender is no longer significant, instead finding two significant interactions with gender. The first shows that, while participants viewed \textsf{creation} of synthetic videos of them saying something as more acceptable than a sexual act, people of marginalized genders were more likely to do so than men ($\text{OR}=10.71$ for men vs. $\text{OR}=3.42$ for marginalized genders, $p=0.027$).

\heading{Participants who are men are more accepting of intimate partners creating synthetic videos 
depicting them.} Secondly, we observe that, holding all other factors constant, men were more likely to rate the \textsf{creation} of synthetic media by an intimate partner more acceptable ($\text{OR}=1.77$ for men vs. $\text{OR}=6.27$ for marginalized genders, $p<0.001$).
Additionally, most participants who described the \textsf{creation} of AIG-NCII in an intimate partnership as being acceptable because it was a compliment or part of their partner's fantasy (as discussed in Section~\ref{sec:results:contextual:fantasy}) were men.
\section{Discussion}
\label{sec:discussion}
Overall, we find that creating, sharing, or seeking AIG-NCII is considered far less acceptable than creating, sharing, or seeking other forms of non-consensually-created synthetic media (RQ1: Section~\ref{sec:results:general:AIGNCII-less-acceptable}). Respondents were more accepting of intimate partners creating synthetic media of them than strangers, including AIG-NCII, but only when their intent in doing so was not to cause harm (RQ2: Section~\ref{sec:results:contextual:intimate-partner-no-harm}). Lack of consent was the most common reason respondents provided for why non-consensual \textsf{creation} of synthetic media, including AIG-NCII, was unacceptable. Our statistical models support this finding: positive attitudes toward sexual consent were inversely correlated with acceptance of non-consensual \textsf{creation}, \textsf{sharing}, or \textsf{seeking\_out} of synthetic media of any kind (RQ3: Section~\ref{sec:results:sexual-consent-content:SCSR}). 
The second most common reason respondents gave for why \textsf{creation} was unacceptable was potential for harm, either reputational damage or bodily violation; conversely, the lack of potential for such harm was the most common reason among those who found \textsf{creation} acceptable.
Men in particular were more accepting of synthetic media \textsf{creation} (RQ4: Section~\ref{sec:results:gender}), especially by intimate partners. 
We hypothesize based on prior literature on perceptions of sexual reputation in the context of defamation law~\cite{baker2011defamation,pruitt2004her,smith1992malice} and participants' open-text responses that this is likely due to differences in perception regarding reputation damage and \textsf{creation} as a form of compliment as well as, from a critical perspective~\cite{CriticalSecurity}, that men may be more accepting of such images if they have more power in a relationship. 
Respondents also expressed attitudes of unacceptability due to moral violations~\cite{giner2018makes}, including feelings of disgust, and privacy violation. 

We focus the remainder of our discussion on implications for addressing the most unacceptable use of AI generative capabilities we find in our study, AIG-NCII, although we note that the implications are relevant to other synthetic media.

\heading{Distributed responsibility and individual deterrence.}
We believe it is important to understand the gap between the unacceptability of creation and sharing and the relative acceptability of searching for, and subsequently viewing, of \aigncii. Based on our results, we hypothesize that one contributing factor to the continued ubiquity of \aigncii is the broad acceptance of or neutrality toward searching for such content. The finding that searching for and viewing \aigncii is perceived as  so acceptable suggests the harms entailed in \aigncii are not fully appreciated by many people. Yet as studies of the experiences of image-based sexual abuse victim-survivors and even legal cases note, viewing is a primary mechanism of harm for NCII: ``there [is] a fresh intrusion of privacy when each additional viewer sees the photograph''~\cite{kerenpaz2023power}.

Past works, although not written in the context of \aigncii, can provide possible explanations for this gap, which we encourage future research to explore in depth. As media scholar Lilie Chouliaraki concludes in her analysis of the viewing of violent imagery in television and online, ``technology closes the moral distance between spectators and sufferers and \ldots\ yet, at the same time, it fictionalizes suffering and leads spectators to indifference''~\cite{chouliaraki2006spectatorship}.
Media scholar Charles Ess~\cite{Ess}, in his foundational work \textit{Digital Media Ethics}, argues that such indifferent online behavior in new media networks is due to ``distributed responsibility,'' which refers to the idea that ethical responsibility for an act is distributed across an interconnected, online networks of actors, rather than being attached solely to a single individual~\cite{Ess,VriesPorsche}. 
Ess contrasts this collective responsibility with the traditional western understanding of ethical responsibility as matter of individual agency. For example, an individual might never steal an album from a physical record store but may illegally download of music from the Internet. In this and many cases, he argues, individuals consider themselves part of an anonymous, undetectable online collective without fear of punishment. 

Thus, a key question for future work is how to combat indifference towards the harm of viewing \aigncii.
Deterrence messaging, such as keyword-based warnings in search engines or advertisements that inform the viewer about the harms of consuming \aigncii, could be used to target individuals' sense of ethical immunity. Emphasizing personal accountability within the collective space could disrupt feelings of distributed responsibly related to \aigncii. Such messaging is currently effectively used to deter viewing of child sexual abuse material~\cite{prichard2022effects} but further research is necessary to find effective approaches to deter \aigncii consumption.

\heading{Harms vs. rights}
When analyzing our data, we observed different classes of arguments for (and against) the unacceptability of AIG-NCII. At the highest level, we saw arguments focused on harms and arguments focused on rights. For example, some argued that creating AIG-NCII was acceptable as long as no harms manifested, e.g., ``It’s not harming me or blackmailing me \ldots\ 
[a]s long as it doesn't get shared I think it’s ok'' (Section~\ref{sec:results:general:harms}): a  harms-based analysis. On the other hand, some argued that creating AIG-NCII was unacceptable, even if never shared, because it was a ``violation of my body'' (Section~\ref{sec:results:general:harms}): a rights-based evaluation.

While prior work on AIG-NCII has primarily focused on harm perceptions~\cite{KuglerDeepfakeAttitudes, FidoCelebrity}, 
these two categories of arguments --- harms-based and rights-based --- align with the vast literature in philosophy and psychology on how different people may center different values in moral decision making, e.g., see~\cite{SecurityEthicsConference} for a  survey aimed at the security and privacy community. Using the terminology from philosophy, those who consider AIG-NCII unacceptable because it can lead to harms are centering a utilitarianistic  (consequentialist) perspective on ethics; those who consider AIG-NCII unacceptable because it violates an individual's rights even if no harms manifest are centering a deontological perspective.

While our findings surfaced a breadth of rights that participants believe are impacted by the creation and possible sharing of \aigncii, we focus on two below: the right to consent, which is baked into the definition of \aigncii, and, given the SOUPS community, the right to privacy.

\heading{\aigncii as a consent violation.}
To our knowledge, ours is the first work to surface qualitative perspectives on consent for \aigncii. 
Our findings (Section~\ref{sec:results:sexual-consent-content}) suggest connections between understandings and norms around consent in different contexts. 
Grounded in the observed relationships among respondents' acceptability ratings, attitudes towards sexual consent, and their free response explanations, we speculate on the potential implications of these context connections: 
First, shaping or enforcing norms around sexual consent, or consent in general, could influence norms and behaviors related to non-consensual synthetic media. Consent education, which involves setting and modeling behavioral norms like asking for consent before interacting with another person's body or space, is one approach to establishing and enforcing norms around consent for all ages in both sexual and non-sexual contexts~\cite{AllAgesConsentEducation, SexualConsentEducation}. 
Second, centering consent as a priority in policies and technical developments around deepfakes is warranted. A growing body of work provides useful frameworks for operationalizing consent in sociotechnical systems~\cite{StrengersHCIConsent, ZytkoHCIConsent, KaregarHCIConsent}.

\heading{\aigncii as a privacy violation.}
Like consent, privacy is a fundamental right. While our survey instrument did not mention privacy at any point, some participants stated that the creation of the synthetic media would violate their privacy. 

The fact that contextual factors such as who created the content and for what purpose influence perceptions of \aigncii acceptability in our study aligns with existing technology privacy theory on contextual norms~\cite{wisniewski2022privacy} and integrity~\cite{NissenbaumPrivacy}, which find that experiences of privacy violation are dependent on contextual factors including what information is being shared, which actors are involved, and the purpose of the information sharing. 
Thus, frameworks of privacy as contextual integrity may be one useful component of future policies about \aigncii.  

At the same time, existing frameworks and technological conceptions of privacy often focus on \textit{data} privacy. Yet, as technological capabilities continue to develop, technologists must increasingly contemplate how to measure and protect a more nebulous privacy right: to representational privacy. Creating \aigncii may involve non-sensitive personal data that becomes sensitive in an \aigncii image. Rather, what is sensitive is a technologically-produced representation of the self made possible using a small amount of personal data (e.g., a photograph of the subject) and a large amount of other people's data (used to train the model that generated the \aigncii). While technical work focusing on detecting sensitive parts of images~\cite{Tonge2020IPP} is valuable and should be continued, protecting representational privacy requires holistic considerations beyond just identifying and redacting sensitive image regions. 

Legal scholars have already begun to wrestle with this issue, highlighting that existing regulation on privacy may not be wholly sufficient to protect sexual autonomy~\cite{CitronSexualPrivacy}. 
Citron proposes the recognition of sexual privacy --- ``the behaviors, expectations, and choices that manage access to and information about the human body, sex, sexuality, gender, and intimate activities''~\cite{CitronSexualPrivacy} --- to provide more holistic protections for subjects of intimate images. What would a similar reformulation from data privacy to representational privacy mean for the technical security and privacy community? Answering this question will require translating notions of self-representation and consent into technical constraints that can govern systems.
\section{Conclusion}
Public familiarity with \aigncii is still low~\cite{UmbachAttitudes}. As more of it is produced~\cite{FlynnIBSADeepfakes} and it becomes easier to produce (e.g., through commercial text-to-video products or ``nudify'' apps~\cite{MurphyNudify, ElsesserNudify}), technological acceptance may increase and attitudes may change~\cite{GranicTAM}. Continued work is needed to track and understand the development of technology for creating and sharing \aigncii as well as the attitudes around it. Our study contributes towards the understanding of attitudes towards non-consensual deepfakes across contexts, including \aigncii, providing insight into the rationales behind people's attitudes as well as the connections between gender, consent, genuine intimate imagery and these attitudes. Addressing \aigncii media requires a multifaceted response blending social science work on norms, legal scholarship, and socio-technical research to detect and prevent creation, sharing and viewing of harmful synthetic media.

\section*{Acknowledgments}
We thank Samuel Dooley for his guidance and feedback on our statistical analysis. We are also grateful to Rosanna Bellini and Sharon Wang for their feedback regarding ethical survey design. Additionally, we appreciate the members of the Security and Privacy Lab at the University of Washington for their insights and brainstorming contributions.
This work was supported in part by NSF Award \#2205171 and the Google PhD Fellowship.

\bibliographystyle{plain}
\bibliography{references}

\appendix
\ifnum\confversion=0
\section{Full Survey Instrument}
\label{ap:full-survey-instrument}
\label{ap:SCS-R}
{\setlength{\parindent}{0pt}
\textcolor{gray}{[Informed Consent]}
This is a survey about generative AI systems that can be used to create realistic-looking, but fake, images, videos, and audio of people. As part of this survey, we discuss different scenarios about this type of media. Be aware that some of these scenarios discuss people using such technology to create synthetic intimate images of you.
The University of Washington’s Human Subjects Division 
reviewed our study and determined that it was exempt from federal human subjects regulation. We do not expect that this survey will put you at more risk than you might encounter in everyday conversations.
\vspace{3pt}

To participate, you must be at least 18 years old and able to complete the survey in English. We expect this survey will take about 8 minutes to complete. If you have any questions about this survey, you may email us at digitalharmstudies@gmail.com.
\vspace{3pt}

I am 18 years or older. $\bigcirc$ Yes $\bigcirc$ No
\vspace{3pt}

I have read and I understand the information above. $\bigcirc$ Yes $\bigcirc$ No
\vspace{3pt}
I want to participate in this survey and continue with the task. $\bigcirc$ Yes $\bigcirc$ No
\vspace{6pt}

\textcolor{gray}{[Background, task description, \& re-consent]}
A generative AI system is a type of artificial intelligence that learns patterns and relationships from a dataset of human-created content. When given a prompt, it uses this learned knowledge to generate new content in a similar style or format. This enables it to autonomously create realistic and coherent outputs, ranging from images to text.
\vspace{3pt}

For example, a generative AI system might learn from a set of photographs taken by people. When prompted to generate an image of the sky, it would use patterns from this data set of photographs to generate the requested content.
\vspace{3pt}

These systems can be used to create realistic-looking, but fake, images and videos of people. For example, generative AI systems can create synthetic intimate images - images of people naked or engaged in sexual acts.
\vspace{3pt}

In this survey, we will show you 3 scenarios about synthetic, AI-generated videos. After presenting each scenario, we will ask you some follow-up questions. Then we will ask a series of questions about non-synthetic intimate imagery and other topics, including topics related to sex and sexuality. We will end the survey by asking general demographic questions.
\vspace{3pt}

I understand what this survey involves and would like to continue. $\bigcirc$ Yes $\bigcirc$ No\\

\textcolor{gray}{[Vignettes (this section is repeated three times with randomized vignettes)]} Imagine that an intimate partner uses generative AI to create a synthetic video of you performing a sexual act for the purpose of sexual pleasure. Assume that you are unaware of the video's creation and existence.
\begin{center}
    \includegraphics[width=0.3\textwidth]{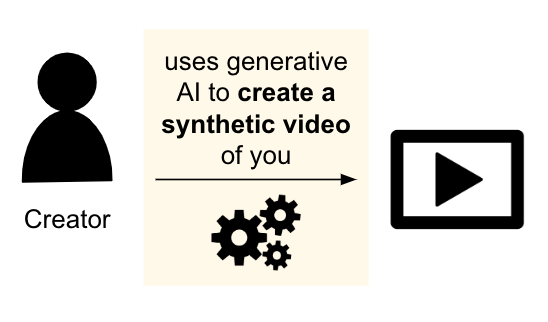}
\end{center}
In my opinion, the creation of this video is: $\bigcirc$ Totally unacceptable $\bigcirc$ Somewhat unacceptable $\bigcirc$ Neutral $\bigcirc$ Somewhat acceptable $\bigcirc$ Totally acceptable $\bigcirc$ Prefer not to answer
\vspace{3pt}

Please write at least 1-2 sentences about why you feel this scenario in particular is (un)acceptable. \rule{1cm}{0.15mm}
\vspace{3pt}

Now, imagine that the creator of this video shares it with other people in a private channel, like a group chat with friends.
\begin{center}
    \includegraphics[width=0.3\textwidth]{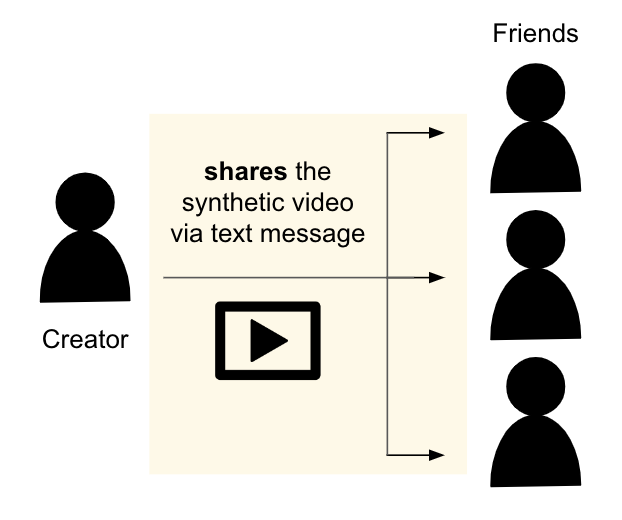}
\end{center}
In my opinion, this sharing of the video is: $\bigcirc$ Totally unacceptable $\bigcirc$ Somewhat unacceptable $\bigcirc$ Neutral $\bigcirc$ Somewhat acceptable $\bigcirc$ Totally acceptable $\bigcirc$ Prefer not to answer
\vspace{3pt}

Now, imagine that one of the people in the group chat, who received the video from the creator through the private channel, reshares the video publicly, like posting it on Reddit.
\begin{center}
    \includegraphics[width=0.4\textwidth]{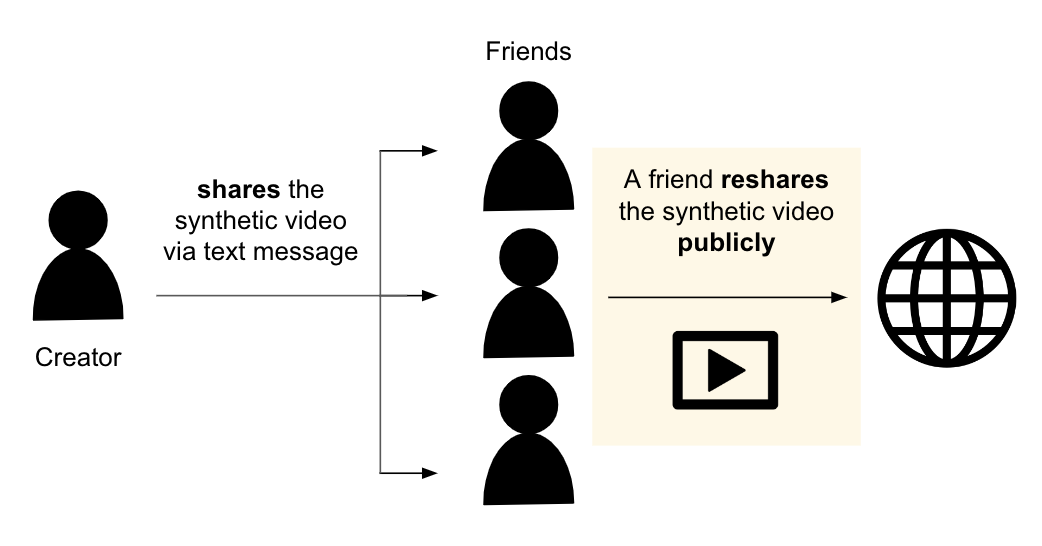}
\end{center}
In my opinion, this resharing of the video is: $\bigcirc$ Totally unacceptable $\bigcirc$ Somewhat unacceptable $\bigcirc$ Neutral $\bigcirc$ Somewhat acceptable $\bigcirc$ Totally acceptable $\bigcirc$ Prefer not to answer
\vspace{3pt}

We are now going to consider an alternative scenario. Imagine the creator of this video shares it in a public format, like posting it on Reddit.
\begin{center}
    \includegraphics[width=0.3\textwidth]{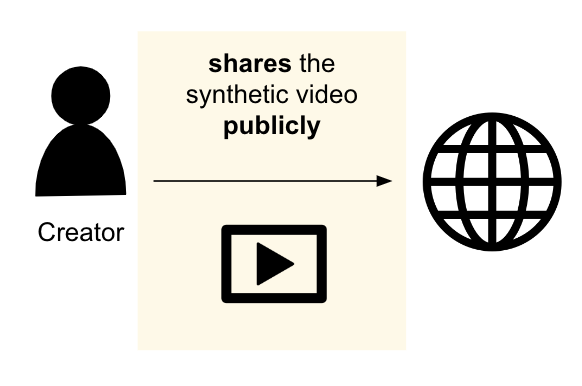}
\end{center}
In my opinion, this sharing of the video is: $\bigcirc$ Totally unacceptable $\bigcirc$ Somewhat unacceptable $\bigcirc$ Neutral $\bigcirc$ Somewhat acceptable $\bigcirc$ Totally acceptable $\bigcirc$ Prefer not to answer
\vspace{3pt}

Now, imagine someone who has not created this video or had it shared with them seeks the video out, like by searching online for a video matching its description.
\begin{center}
    \includegraphics[width=0.3\textwidth]{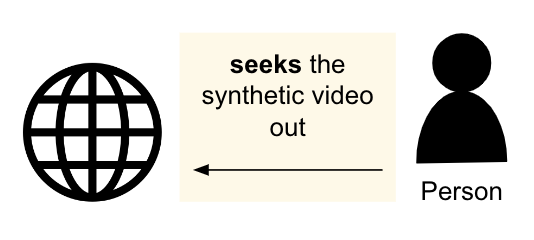}
\end{center}
In my opinion, the seeking out of this video is: $\bigcirc$ Totally unacceptable $\bigcirc$ Somewhat unacceptable $\bigcirc$ Neutral $\bigcirc$ Somewhat acceptable $\bigcirc$ Totally acceptable $\bigcirc$ Prefer not to answer
\vspace{3pt}

People take many different approaches to handling online content and experiences. If the video was shared in a public format, which of the following, if any, would you do: $\square$ Report this content to the platform for removal $\square$ Reach out to a helpline or support service organization for assistance getting the content removed $\square$ Reach out to a lawyer $\square$ Reach out to the police $\square$ Reach out to friends or family $\square$ Other \rule{1cm}{0.15mm} $\square$ None of the above $\square$ Prefer not to answer
\vspace{3pt}

This marks the end of questions about this scenario.
\vspace{6pt}

\textcolor{gray}{[Authentic intimate imagery questions]} The following questions are about non-synthetic, intimate imagery.
\vspace{3pt}

Imagine two people in an intimate relationship (e.g., dating, married) send nude media (images or videos) of themselves to each other.
\begin{center}
    \includegraphics[width=0.3\textwidth]{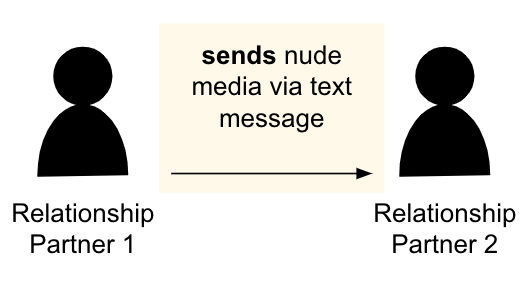}
\end{center}
In my opinion this is: $\bigcirc$ Totally unacceptable $\bigcirc$ Somewhat unacceptable $\bigcirc$ Neutral $\bigcirc$ Somewhat acceptable $\bigcirc$ Totally acceptable $\bigcirc$ Prefer not to answer
\vspace{3pt}

Now, imagine that the intended recipient of this media shares it with other people in a private channel, like a group chat with friends, without informing the original sender.
\begin{center}
    \includegraphics[width=0.4\textwidth]{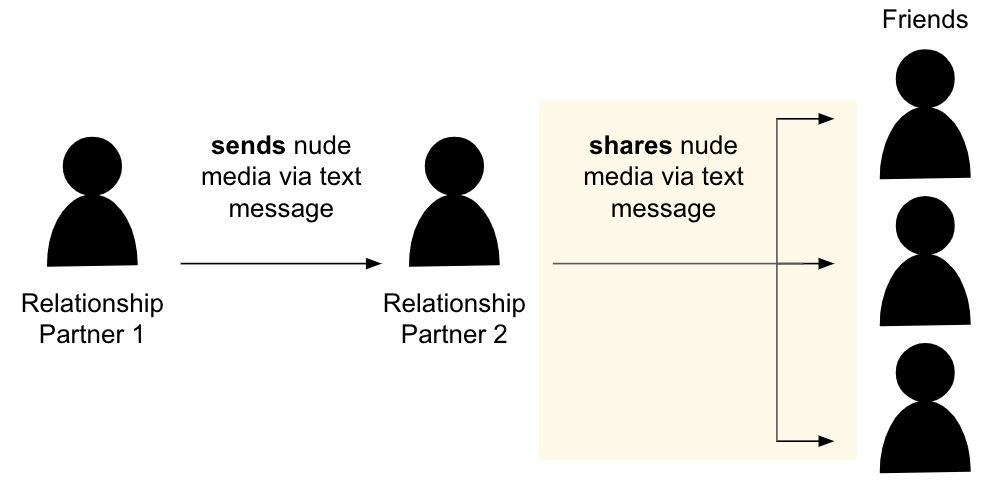}
\end{center}
In my opinion, this sharing of the media is: $\bigcirc$ Totally unacceptable $\bigcirc$ Somewhat unacceptable $\bigcirc$ Neutral $\bigcirc$ Somewhat acceptable $\bigcirc$ Totally acceptable $\bigcirc$ Prefer not to answer
\vspace{3pt}

Now, imagine that one of the people in the group chat, who received the media from the intended recipient through the private channel, reshares the media publicly, like posting it on Reddit.
\begin{center}
    \includegraphics[width=0.4\textwidth]{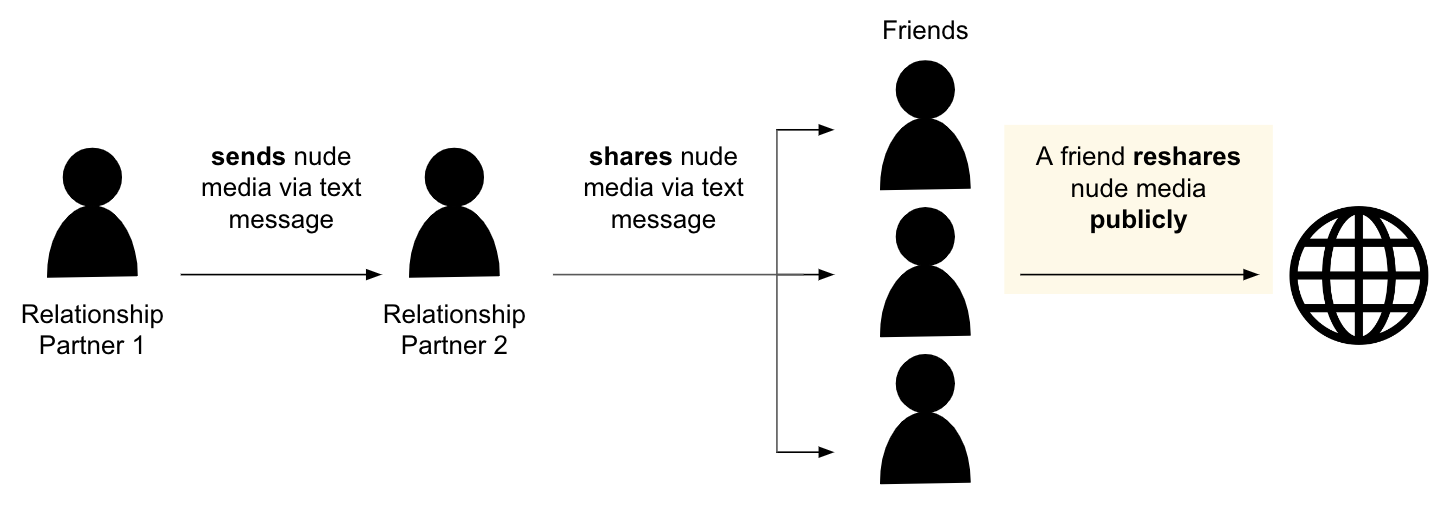}
\end{center}
In my opinion, this resharing of the video is: $\bigcirc$ Totally unacceptable $\bigcirc$ Somewhat unacceptable $\bigcirc$ Neutral $\bigcirc$ Somewhat acceptable $\bigcirc$ Totally acceptable $\bigcirc$ Prefer not to answer
\vspace{3pt}

We are now going to consider an alternative scenario. Imagine someone in the relationship shares nude media that they received from the other person in a public format, like posting it on Reddit.
\begin{center}
    \includegraphics[width=0.4\textwidth]{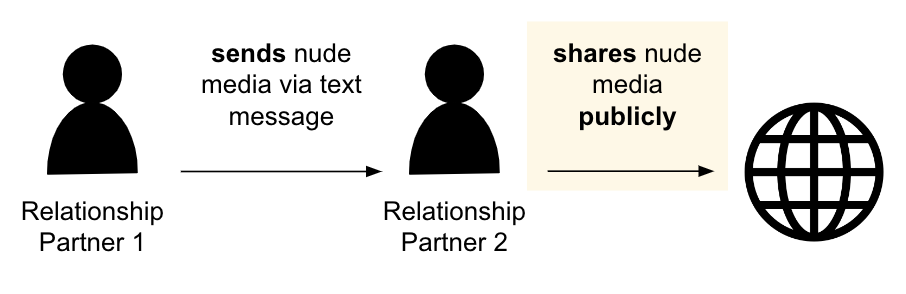}
\end{center}
In my opinion, this sharing of the media is: $\bigcirc$ Totally unacceptable $\bigcirc$ Somewhat unacceptable $\bigcirc$ Neutral $\bigcirc$ Somewhat acceptable $\bigcirc$ Totally acceptable $\bigcirc$ Prefer not to answer
\vspace{3pt}

Now, imagine someone outside of the relationship seeks the video out, like by searching online for media matching its description.
\begin{center}
    \includegraphics[width=0.3\textwidth]{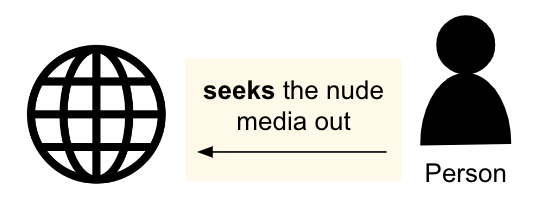}
\end{center}
In my opinion, the seeking out of this video is: $\bigcirc$ Totally unacceptable $\bigcirc$ Somewhat unacceptable $\bigcirc$ Neutral $\bigcirc$ Somewhat acceptable $\bigcirc$ Totally acceptable $\bigcirc$ Prefer not to answer
\vspace{3pt}

People take many different approaches to handling online content and experiences. If you were the subject of this media and it was shared in a public format, which of the following, if any would you do: $\square$ Report this content to the platform for removal $\square$ Reach out to a helpline or support service organization for assistance getting the content removed $\square$ Reach out to a lawyer $\square$ Reach out to the police $\square$ Reach out to friends or family $\square$ Other \rule{1cm}{0.15mm} $\square$ None of the above $\square$ Prefer not to answer
\vspace{3pt}

Indicate your agreement with the following: “If a nude photo or video is shared with anyone, it will eventually be seen by people who were not the intended recipients.” $\bigcirc$ Strongly disagree $\bigcirc$ disagree $\bigcirc$ Neither agree nor disagree $\bigcirc$ Agree $\bigcirc$ Strongly agree $\bigcirc$ Prefer not to answer
\vspace{6pt}

\textcolor{gray}{[Deepfake community attitude agreement (these statements were sourced from a study of online deepfake communities ~\cite{TimmermanDeepfakeCommunity} and were not used in analysis)]} Realistic-looking fake porn can now be created using generative AI to swap the faces of pornographic actors with other people’s faces, so that it looks like they’re in a porn video. This is sometimes referred to as `deepfake pornography' as it looks very realistic and can be hard to recognize as fake or digitally created.
\vspace{3pt}

Imagine reading the following statement online: “Using generative AI to create pornography another person without their consent is acceptable.” Indicate your agreement with this statement: $\bigcirc$ Strongly disagree $\bigcirc$ disagree $\bigcirc$ Neither agree nor disagree $\bigcirc$ Agree $\bigcirc$ Strongly agree $\bigcirc$ Prefer not to answer
\vspace{3pt}

Imagine reading the following statement online: “A person should be able to create deepfake pornography of anyone without their consent.” Indicate your agreement with this statement: $\bigcirc$ Strongly disagree $\bigcirc$ disagree $\bigcirc$ Neither agree nor disagree $\bigcirc$ Agree $\bigcirc$ Strongly agree $\bigcirc$ Prefer not to answer
\vspace{3pt}

Imagine reading the following statement online: “Creating deepfake pornography of someone whose authentic nude media is publicly available does not cause additional harm.” Indicate your agreement with this statement: $\bigcirc$ Strongly disagree $\bigcirc$ disagree $\bigcirc$ Neither agree nor disagree $\bigcirc$ Agree $\bigcirc$ Strongly agree $\bigcirc$ Prefer not to answer
\vspace{3pt}

Imagine reading the following statement online: “Violence is an act committed against a person that results in bodily harm, physical abuse is violence for example, sexual assault is violence, a video is not violence. Even a fake video of fake violence is not violence.” Indicate your agreement with this statement: $\bigcirc$ Strongly disagree $\bigcirc$ disagree $\bigcirc$ Neither agree nor disagree $\bigcirc$ Agree $\bigcirc$ Strongly agree $\bigcirc$ Prefer not to answer
\vspace{3pt}

Imagine reading the following statement online: “Even clay modeling and cave paintings were used for porn. There is always resistance to novelty. But even if there might have been a temporary ban on those Arts, they still went on. Same here. We just gotta show some diversity.” Indicate your agreement with this statement: $\bigcirc$ Strongly disagree $\bigcirc$ disagree $\bigcirc$ Neither agree nor disagree $\bigcirc$ Agree $\bigcirc$ Strongly agree $\bigcirc$ Prefer not to answer
\vspace{3pt}

Imagine reading the following statement online: “Technically, this is a cropping of one face onto another body. Has been done since decades. But for pictures. Now the pictures are moving.” Indicate your agreement with this statement: $\bigcirc$ Strongly disagree $\bigcirc$ disagree $\bigcirc$ Neither agree nor disagree $\bigcirc$ Agree $\bigcirc$ Strongly agree $\bigcirc$ Prefer not to answer
\vspace{3pt}

Imagine reading the following statement online: “I’d be thrilled that someone found me attractive enough to be worth making a deepfake of lol!” Indicate your agreement with this statement: $\bigcirc$ Strongly disagree $\bigcirc$ disagree $\bigcirc$ Neither agree nor disagree $\bigcirc$ Agree $\bigcirc$ Strongly agree $\bigcirc$ Prefer not to answer
\vspace{6pt}

\textcolor{gray}{[Sexual Consent Scale - Revised Subscale 2: Positive attitudes towards establishing sexual consent]}\\
\textcolor{gray}{Response options: $\bigcirc$ Strongly disagree $\bigcirc$ Disagree $\bigcirc$ Somewhat disagree $\bigcirc$ Neither agree nor disagree $\bigcirc$ Somewhat agree $\bigcirc$ Agree $\bigcirc$ Strongly agree} 
\begin{enumerate}[itemsep=0ex]
    \item I feel that sexual consent should always be obtained before the start of any sexual activity
    \item I believe that asking for sexual consent is in my best interest because it reduces any misinterpretations that might arise
    \item I think it is equally important to obtain sexual consent in all relationships regardless of whether or not they have had sex before
    \item I feel that verbally asking for sexual consent should occur before proceeding with any sexual activity
    \item When initiating sexual activity, I believe that one should always assume they do not have sexual consent
    \item I believe that it is just as necessary to obtain consent for genital fondling as it is for sexual intercourse
    \item Most people that I care about feel that asking for sexual consent is something I should do
    \item I think that consent should be asked before any kind of sexual behavior, including kissing or petting
    \item I feel it is the responsibility of both partners to make sure sexual consent is established before sexual activity begins
    \item Before making sexual advances, I think that one should assume ``no'' until there is a clear indication to proceed
    \item Not asking for sexual consent some of the time is ok
\end{enumerate}
\textcolor{gray}{Scoring instructions: Strongly disagree = 1, Disagree = 2, Somewhat disagree = 3, Neither agree nor disagree = 4, Somewhat agree = 5, Agree = 6, Strongly agree = 7. Reverse item 11 (1 = 7, 2 = 6, 3 = 5, 4 = 4, 5 = 3, 6 = 2, 7 = 1) and average all items.} \\

\textcolor{gray}{[Sexual Consent Scale - Revised Subscale 4: Sexual consent norms]}\\
\textcolor{gray}{Response options: $\bigcirc$ Strongly disagree $\bigcirc$ Disagree $\bigcirc$ Somewhat disagree $\bigcirc$ Neither agree nor disagree $\bigcirc$ Somewhat agree $\bigcirc$ Agree $\bigcirc$ Strongly agree}
\begin{enumerate}[itemsep=0ex]
    \item I think that obtaining sexual consent is more necessary in a new relationship than in a committed relationship
    \item I think that obtaining sexual consent is more necessary in a casual sexual encounter than in a committed relationship
    \item I believe that the need for asking for sexual consent decreases as the length of an intimate relationship increases
    \item I believe it is enough to ask for consent at the beginning of a sexual encounter
    \item I believe that sexual intercourse it the only sexual activity that requires explicit verbal consent
    \item I believe that partners are less likely to ask for sexual consent the longer they are in a relationship
    \item If consent for sexual intercourse is established, petting and fondling can be assumed
\end{enumerate}
\textcolor{gray}{Scoring instructions: Strongly disagree = 1, Disagree = 2, Somewhat disagree = 3, Neither agree nor disagree = 4, Somewhat agree = 5, Agree = 6, Strongly agree = 7. Average all items.}
\vspace{6pt}

\textcolor{gray}{[Demographics]} What is your age? \rule{1cm}{0.15mm}
\vspace{3pt}

What is your gender? \rule{1cm}{0.15mm}
\vspace{3pt}

In politics today, do you consider yourself a Republican, a Democrat, an Independent, or something else? $\bigcirc$ Republican $\bigcirc$ Democrat $\bigcirc$ Independent $\bigcirc$ Don't know $\bigcirc$ Refuse to answer $\bigcirc$ Other
\vspace{3pt}

\textcolor{gray}{(If answered `Independent,' `Don't know,' `Refuse to answer,' or `Other')} As of today, do you lean more to the Republican Party or the Democratic Party? $\bigcirc$ Republican $\bigcirc$ Democrat $\bigcirc$ Refuse to answer
\vspace{3pt}

\textcolor{gray}{[Attention check]} What would you like to see elected leaders in Washington get done during the next few years? Please give as much detail as you can. \rule{1cm}{0.15mm}
\vspace{6pt}

\textcolor{gray}{[Debrief]} If you or someone you know experiences or has experienced non-consensual intimate image abuse, support is available. Visit https://StopNCII.org/ and https://cybercivilrights.org/ccri-safety-center/ for comprehensive resources and information. Additionally, you can contact the National Domestic Violence Hotline [1(800)799-SAFE (7233)] for confidential assistance and guidance. Select "Next Page" to complete the survey and be redirected to Prolific.
}

\fi

\section{Participant demographics}
\label{ap:demographics}
Participants' gender, age, and political orientation is presented in Table~\ref{ap:demographics}.
\begin{table*}[]
  \centering
  \begin{tabular}{lll}
    \toprule
    Gender & Age & Political Orientation \\
    \midrule
    \begin{tabular}{@{}lc@{}}
      Woman & 49.5\% \\
      Man & 47.6\% \\
      Non-binary & 1.9\% \\
      Agender & 0.6\% \\
      Prefer not to say & 0.3\%
    \end{tabular}
    &
    \begin{tabular}{@{}lc@{}}
      18-24 & 17.8\% \\
      25-34 & 33.0\% \\
      35-44 & 24.4\% \\
      45-54 & 13.3\% \\
      55-64 & 7.9\% \\
      65+ & 2.9\% \\
      Prefer not to say & 0.6\% 
    \end{tabular}
    &
    \begin{tabular}{@{}lc@{}}
      Democrat & 48.6\% \\
      Republican & 16.2\% \\
      Leans Democrat & 18.4\% \\
      Leans Republican & 8.9\% \\
      Refuse to answer & 7.9\% \\
    \end{tabular} \\
    \bottomrule
  \end{tabular}
  \caption{Breakdown of participant demographics by gender, age, and political orientation.}
  \label{tab:participantDemographics}
\end{table*}

\section{Media action regression results}
\label{ap:metaaction-regression}
\ifnum\confversion=1

\begin{table}[]
\renewcommand{\arraystretch}{1.2}
\centering
\footnotesize
\begin{tabular}{m{0.3cm}m{4.1cm}m{3cm}}
\toprule
& & OR; Confidence Interval \\
\midrule
\multirow{4}[0]{*}{\rotatebox[origin=c]{90}{\textbf{Intercepts\hspace{5pt}}}}
& \vspace{2pt} \cellcolor{yellow!30} \shortstack[l]{Totally unacceptable | \\\;\;Somewhat unacceptable} &  \cellcolor{yellow!30}2.42; [1.89, 3.1]*** \\
& \cellcolor{yellow!30} Somewhat unacceptable | Neutral & \cellcolor{yellow!30}7.41; [5.73, 9.59]*** \\
& \cellcolor{yellow!30} Neutral | Somewhat acceptable & \cellcolor{yellow!30}28.59; [21.65, 37.76]*** \\
& \vspace{2pt} \cellcolor{yellow!30}\shortstack[l]{Somewhat acceptable | \\\;\;Totally acceptable} & \cellcolor{yellow!30}89.53;[65.78, 121.85]*** \\
\midrule
\multirow{4}[0]{*}{\rotatebox[origin=c]{90}{\textbf{\shortstack{Content\\Action}}}}
& \cellcolor{yellow!30} \textsf{private\_sharing} & \cellcolor{yellow!30}0.47; [0.37, 0.58]*** \\
& \cellcolor{yellow!30} \textsf{public\_sharing} & \cellcolor{yellow!30}0.26; [0.21, 0.33]*** \\
& \cellcolor{yellow!30} \textsf{resharing} & \cellcolor{yellow!30}0.42; [0.33, 0.52]*** \\
& \cellcolor{yellow!30} \textsf{seeking\_out} & \cellcolor{yellow!30}5.43; [4.45, 6.62]*** \\
\bottomrule
\end{tabular}
\caption{Results from a single regression exploring the relationship between acceptability (first row, intercepts) and action being preformed with the synthetic media (second row, content action). Reference level of content action is \textsf{creation}. Significance of OR: $p<0.001$ = \colorbox{yellow!30}{***}.}
\label{tab:metaactionRegression}
\end{table}

\else

\begin{table*}[]
\renewcommand{\arraystretch}{1.2}
\centering
\footnotesize
\begin{tabular}{m{0.3cm}m{4.1cm}m{3cm}}
\toprule
& & OR; Confidence Interval \\
\midrule
\multirow{4}[0]{*}{\rotatebox[origin=c]{90}{\textbf{Intercepts\hspace{5pt}}}}
& \vspace{2pt} \cellcolor{yellow!30} \shortstack[l]{Totally unacceptable | \\\;\;Somewhat unacceptable} &  \cellcolor{yellow!30}2.42; [1.89, 3.1]*** \\
& \cellcolor{yellow!30} Somewhat unacceptable | Neutral & \cellcolor{yellow!30}7.41; [5.73, 9.59]*** \\
& \cellcolor{yellow!30} Neutral | Somewhat acceptable & \cellcolor{yellow!30}28.59; [21.65, 37.76]*** \\
& \vspace{2pt} \cellcolor{yellow!30}\shortstack[l]{Somewhat acceptable | \\\;\;Totally acceptable} & \cellcolor{yellow!30}89.53;[65.78, 121.85]*** \\
\midrule
\multirow{4}[0]{*}{\rotatebox[origin=c]{90}{\textbf{\shortstack{Content\\Action}}}}
& \cellcolor{yellow!30} \textsf{private\_sharing} & \cellcolor{yellow!30}0.47; [0.37, 0.58]*** \\
& \cellcolor{yellow!30} \textsf{public\_sharing} & \cellcolor{yellow!30}0.26; [0.21, 0.33]*** \\
& \cellcolor{yellow!30} \textsf{resharing} & \cellcolor{yellow!30}0.42; [0.33, 0.52]*** \\
& \cellcolor{yellow!30} \textsf{seeking\_out} & \cellcolor{yellow!30}5.43; [4.45, 6.62]*** \\
\bottomrule
\end{tabular}
\caption{Results from a single regression exploring the relationship between acceptability (first row, intercepts) and action being preformed with the synthetic media (second row, content action). Reference level of content action is \textsf{creation}. Significance of OR: $p<0.001$ = \colorbox{yellow!30}{***}.}
\label{tab:metaactionRegression}
\end{table*}
\fi
Results for the regression are presented in Table~\ref{tab:metaactionRegression}.  

\section{Qualitative Codebook}
\label{ap:qualitative-codebook}
The codebooks from qualitatively analyzing explanations for why the creation of the synthetic video in each vignettes is either acceptable or unacceptable. Codes were not mutually exclusive.

\heading{Rationales for acceptability:}
\begin{itemize}[itemsep=0.1em]
    \item \textbf{No Harm}: Will not cause harm
    \item \textbf{Relationship}: Trust in an intimate partner
    \item \textbf{Indifference}: No impact; `I don't care'
    \item \textbf{Compliment}: Indicates attraction
    \item \textbf{Fantasy}: Indulges fantasy
    \item \textbf{Pro-Tech}: Technology and AI are interesting
\end{itemize}

\heading{Rationales for unacceptability:}
\begin{itemize}[itemsep=0.1em]
    \item \textbf{Consent}: Absence of consent or permission
    \item \textbf{Awareness}: Lack of awareness about video's creation and existence
    \item \textbf{Dislike}: Elicits negative feelings; The video is `weird,' `creepy,' `disgusting,' `uncomfortable,' etc.
    \item \textbf{Harm}: Creates or could create harm
    \item \textbf{Ethics}: Violation of ethics, morality, or law; The video is `wrong'
    \item \textbf{Privacy}: Violation of privacy
    \item \textbf{Fake}: Fake nature, inauthentic
    \item \textbf{Stranger}: Created by a stranger
    \item \textbf{Relationship}: Violation of trust in an intimate partner
\end{itemize}

\section{Additional Models}
\label{ap:additionalModels}
Regression analyses conducted with gender categorized into `men' and `women,' rather than `men' and `marginalized genders.' Eight participants who identified outside of the gender binary or did not disclose their gender were excluded from these analyses. 
For the results with gender bucketed into `men' and `women,' Table~\ref{tab:full-regression-men-women} corresponds to Table~\ref{tab:FullRegression}, Table~\ref{tab:contextual-interaction-men-women} to Table~\ref{tab:contextualInteractionRegressionTable}, and Table~\ref{tab:gender-interaction-men-women} to Table~\ref{tab:genderInteractionRegressionTable}.
\ifnum\confversion=1
\\\\\\\\\\\\\\
\fi

\begin{table*}[t]
    \centering
    \footnotesize
    \renewcommand{\arraystretch}{2.4}
    \newcolumntype{Y}{>{\centering\arraybackslash}X}
    \newcolumntype{M}[1]{>{\centering\arraybackslash}m{#1}} 
    \begin{tabularx}{\textwidth}{M{0.3cm} M{6cm} *{5}{Y}}
    \toprule
        & & \textsf{creation} 
        & \textsf{private\_sharing}  
        & \textsf{public\_sharing}
        & \textsf{resharing}
        & \textsf{seeking\_out}  \\
    \midrule
    \multirow{4}{*}{\rotatebox[origin=c]{90}{\textbf{Intercepts\hspace{5pt}}}} 
        & Totally unacceptable | Somewhat unacceptable
        & \shortstack{5.86 \\ $[0.47, 73.58]$ } 
        & \shortstack{1.48 \\ $[0.07, 29.23]$ } 
        & \shortstack{22.30 \\ $[0.64, 779.31]$ } 
        & \shortstack{1.36 \\ $[0.05, 24.06]$ } 
        & \cellcolor{yellow!30}\shortstack{0.03* \\ $[0, 1]$ } \\
        & Somewhat unacceptable | Neutral
        & \cellcolor{yellow!30}\shortstack{34.07** \\ $[2.66, 436.98]$ } 
        & \shortstack{7.09 \\ $[0.35, 141.55]$ } 
        & \cellcolor{yellow!30}\shortstack{91.48* \\ $[2.54, 3297.44]$ } 
        & \shortstack{6.03 \\ $[0.24, 152.77]$ } 
        & \shortstack{0.18 \\ $[0.01, 5.37]$ } \\
        & Neutral | Somewhat acceptable
        & \cellcolor{yellow!30}\shortstack{114.61*** \\ $[8.77, 1497.84]$ } 
        & \cellcolor{yellow!30}\shortstack{22.42* \\ $[1.11, 451.16]$ } 
        & \cellcolor{yellow!30}\shortstack{216.15** \\ $[5.89, 7933.25]$ } 
        & \shortstack{19.72 \\ $[0.77, 503.89]$ } 
        & \shortstack{6.08 \\ $[0.21, 174.27]$ } \\
        & Somewhat acceptable | Totally acceptable 
        & \cellcolor{yellow!30}\shortstack{426.21*** \\ $[31.71, 5728.35]$ } 
        & \cellcolor{yellow!30}\shortstack{101.08** \\ $[4.89, 2089.5]$ } 
        & \cellcolor{yellow!30}\shortstack{1104.00*** \\ $[28.58, 42643.48]$ } 
        & \cellcolor{yellow!30}\shortstack{91.20** \\ $[3.47, 2398.75]$ } 
        & \cellcolor{yellow!30}\shortstack{28.93* \\ $[1, 835.39]$ } \\
    \midrule
    \multirow{5}{*}{\rotatebox[origin=c]{90}{\textbf{Controlled IVs\hspace{5pt}}}} 
        & \textsf{creator} (Intimate partner) 
        & \cellcolor{yellow!30}\shortstack{3.29*** \\ $[2.25, 4.79]$ } 
        & \cellcolor{yellow!30}\shortstack{1.71* \\ $[1.13, 2.58]$ } 
        & \shortstack{1.45 \\ $[0.89, 2.37]$ } 
        & \shortstack{1.04 \\ $[0.67, 1.6]$ } 
        & \shortstack{1.13 \\ $[0.81, 1.57]$ } \\
        & \textsf{action} (Sport) 
        & \cellcolor{yellow!30}\shortstack{12.96*** \\ $[7.69, 21.85]$ } 
        & \cellcolor{yellow!30}\shortstack{33.43*** \\ $[16.13, 69.26]$ } 
        & \cellcolor{yellow!30}\shortstack{64.11*** \\ $[22.29, 184.37]$ } 
        & \cellcolor{yellow!30}\shortstack{31.32*** \\ $[14.57, 67.29]$ } 
        & \cellcolor{yellow!30}\shortstack{7.22*** \\ $[4.68, 11.15]$ } \\
        & \textsf{action} (Saying something)  
        & \cellcolor{yellow!30}\shortstack{5.48*** \\ $[3.29, 9.14]$ } 
        & \cellcolor{yellow!30}\shortstack{10.58*** \\ $[5.24, 21.34]$ } 
        & \cellcolor{yellow!30}\shortstack{19.33*** \\ $[6.92, 53.97]$ } 
        & \cellcolor{yellow!30}\shortstack{12.26*** \\ $[5.79, 25.95]$ } 
        & \cellcolor{yellow!30}\shortstack{3.39*** \\ $[2.19, 5.23]$ } \\
        & \textsf{intent} (Entertainment) 
        & \cellcolor{yellow!30}\shortstack{19.27*** \\ $[11.17, 33.24]$ } 
        & \cellcolor{yellow!30}\shortstack{12.04*** \\ $[6.85, 21.17]$ } 
        & \cellcolor{yellow!30}\shortstack{10.89*** \\ $[5.59, 21.19]$ } 
        & \cellcolor{yellow!30}\shortstack{5.80*** \\ $[3.132, 10.15]$ } 
        & \cellcolor{yellow!30}\shortstack{4.83*** \\ $[3.16, 7.4]$ } \\
        & \textsf{intent} (Sexual pleasure)
        & \cellcolor{yellow!30}\shortstack{7.51*** \\ $[4.45, 12.68]$ } 
        & \shortstack{1.42 \\ $[0.81, 2.5]$ } 
        & \shortstack{1.17 \\ $[0.59, 2.32]$ } 
        & \shortstack{0.92 \\ $[0.52, 1.64]$ } 
        & \shortstack{1.29 \\ $[0.86, 1.92]$ } \\
    \midrule
    \multirow{4}{*}{\rotatebox[origin=c]{90}{\textbf{Uncontrolled IVs\hspace{5pt}}}} 
        & Gender (Man) 
        & \cellcolor{yellow!30}\shortstack{2.35** \\ $[1.39, 3.99]$ } 
        & \cellcolor{yellow!30}\shortstack{2.06* \\ $[1.18, 3.61]$ } 
        & \shortstack{1.64 \\ $[0.82, 3.27]$ } 
        & \shortstack{1.38 \\ $[0.73, 2.63]$ }
        & \shortstack{1.56 \\ $[0.78, 3.12]$ } \\
        & \gii \& NDII attitudes (Unacceptable) 
        & \cellcolor{yellow!30}\shortstack{0.21* \\ $[0.05, 0.85]$ } 
        & \cellcolor{yellow!30}\shortstack{0.08** \\ $[0.01, 0.4]$ } 
        & \cellcolor{yellow!30}\shortstack{0.09** \\ $[0.02, 0.43]$ } 
        & \cellcolor{yellow!30}\shortstack{0.01*** \\ $[0, 0.05]$ } 
        & \cellcolor{yellow!30}\shortstack{0.01*** \\ $[0.01, 0.03]$ } \\
        & SCS-R2
        & \cellcolor{yellow!30}\shortstack{0.53*** \\ $[0.39, 0.72]$ } 
        & \cellcolor{yellow!30}\shortstack{0.56*** \\ $[0.4, 0.77]$ } 
        & \cellcolor{yellow!30}\shortstack{0.64* \\ $[0.43, 0.96]$ } 
        & \shortstack{0.76 \\ $[0.52, 1.11]$ }
        & \shortstack{0.73 \\ $[0.49, 1.11]$ } \\
        & SCS-R4
        & \shortstack{1.10 \\ $[0.85, 1.43]$ } 
        & \shortstack{1.14 \\ $[0.87, 1.5]$ } 
        & \shortstack{1.25 \\ $[0.89, 1.76]$ } 
        & \shortstack{1.33 \\ $[0.97, 1.83]$ }
        & \shortstack{1.17 \\ $[0.84, 1.65]$ } \\
    \bottomrule
    \end{tabularx}
    \caption{\small{Results from regressions exploring the relationship between scenario acceptability (first row, intercepts), contextual factors (second row, controlled IVs), and personal factors (third row, uncontrolled IVs). Each column represents the output of one regression model. Numeric cells list the odds ratio (OR) and the 95\% confidence interval.  Reference levels: \textsf{creator} (stranger), \textsf{action} (sexual act), \textsf{intent} (harm), gender (woman), \gii \& NDII attitudes (acceptable). Significance of OR: $p<0.05$ = \colorbox{yellow!30}{*}, $p<0.01$ = \colorbox{yellow!30}{**}, and $p<0.001$ = \colorbox{yellow!30}{***}.}} 
    \label{tab:full-regression-men-women}
\end{table*}
\begin{table}[t]
\renewcommand{\arraystretch}{1.2}
\centering
\footnotesize
\begin{tabular}{m{0.3cm}m{4.1cm}m{3cm}}
\toprule
& & OR; Confidence Interval \\
\midrule
\multirow{4}[0]{*}{\rotatebox[origin=c]{90}{\textbf{Intercepts\hspace{5pt}}}}
& \vspace{2pt} \shortstack[l]{Totally unacceptable | \\\;\;Somewhat unacceptable} & 5.83; [0.36, 80.73] \\
& \cellcolor{yellow!30}  Somewhat unacceptable | Neutral & \cellcolor{yellow!30}33.60; \cellcolor{yellow!30}[2.2, 513.89]* \\
& \cellcolor{yellow!30} Neutral | Somewhat acceptable & \cellcolor{yellow!30}118.44; \cellcolor{yellow!30}[7.63, 1838.84]*** \\
& \vspace{2pt} \cellcolor{yellow!30}\shortstack[l]{Somewhat acceptable | \\\;\;Totally acceptable} & \cellcolor{yellow!30}462.15; [29, 7364.66]*** \\
\midrule
\multirow{4}[0]{*}{\rotatebox[origin=c]{90}{\textbf{Controlled IVs\hspace{3pt}}}}
& \cellcolor{yellow!30} \textsf{creator} (Intimate partner) & \cellcolor{yellow!30}1.76; [1.03, 3.01]* \\
& \cellcolor{yellow!30} \textsf{action} (Sport) & \cellcolor{yellow!30}16.03;[7.31, 35.16]*** \\
& \cellcolor{yellow!30} \textsf{action} (Saying something) & \cellcolor{yellow!30}10.95; [4.92, 24.37]*** \\
& \cellcolor{yellow!30} \textsf{intent} (Entertainment) & \cellcolor{yellow!30}20.64; [9.49, 44.9]*** \\
& \cellcolor{yellow!30} \textsf{intent} (Sexual pleasure) & \cellcolor{yellow!30}4.90; [2.29, 10.47]*** \\
\midrule
\multirow{4}[0]{*}{\rotatebox[origin=c]{90}{\textbf{\shortstack{Uncontrolled\\IVs}}}}
& Gender (man) & 1.44; [0.39, 5.27] \\
& \cellcolor{yellow!30} \gii \& NDII attitudes (Unacceptable) & \cellcolor{yellow!30}0.2; [0.05, 0.85]* \\
& \cellcolor{yellow!30} SCS-R2 & \cellcolor{yellow!30}0.52; [0.38, 0.72]*** \\
& SCS-R4 & 1.13; [0.86, 1.47] \\
\midrule
\multirow{4}[0]{*}{\rotatebox[origin=c]{90}{\textbf{Interaction Terms\hspace{24pt}}}}
& \shortstack[l]{\textsf{action} (Sport) \& \\\;Gender (Man)} & 0.84; [0.32, 2.23] \\
& \vspace{2pt} \cellcolor{yellow!30} \shortstack[l]{\textsf{action} (Saying something) \& \\\;Gender (Man)} & \cellcolor{yellow!30} 0.31; [0.11, 0.86]* \\
& \vspace{2pt} \shortstack[l]{\textsf{intent} (Entertainment) \& \\\;Gender (Man)} & 0.97; [0.37, 2.59] \\
& \vspace{2pt} \shortstack[l]{\textsf{intent} (Sexual pleasure) \& \\\;Gender (Man)} & 2.38; [0.86, 6.53] \\
& \vspace{2pt} \cellcolor{yellow!30} \shortstack[l]{\textsf{creator} (Intimate partner) \& \\\;Gender (Man)}& \cellcolor{yellow!30}3.58; [1.7, 7.56]*** \\
\bottomrule
\end{tabular}
\caption{Results from a single regression exploring the relationship between scenario acceptability for \textsf{creation} (first row, intercepts), contextual factors (second row, controlled IVs), personal factors (third row, uncontrolled IVs), and interactions between gender and contextual factors (third row, interaction terms). Reference levels: \textsf{creator} (stranger), \textsf{action} (sexual act), \textsf{intent} (harm), gender (woman), \gii \& NDII attitudes (acceptable). Significance of OR: $p<0.05$ = \colorbox{yellow!30}{*}, $p<0.01$ = \colorbox{yellow!30}{**}, and $p<0.001$ = \colorbox{yellow!30}{***}.}
\label{tab:gender-interaction-men-women}
\end{table}
\begin{table}[t]
\renewcommand{\arraystretch}{1.2}
\centering
\footnotesize
\begin{tabular}{m{0.3cm}m{4.1cm}m{3cm}}
\toprule
& & OR; Confidence Interval \\
\midrule
\multirow{4}[0]{*}{\rotatebox[origin=c]{90}{\textbf{Intercepts\hspace{5pt}}}}
& \vspace{2pt} \shortstack[l]{Totally unacceptable | \\\;\;Somewhat unacceptable} & 8.90; [0.49, 163.28] \\
& \cellcolor{yellow!30}  Somewhat unacceptable | Neutral & \cellcolor{yellow!30}57.23; [3.07, 1067.96]** \\
& \cellcolor{yellow!30} Neutral | Somewhat acceptable & \cellcolor{yellow!30}204.33; [10.78, 3872.9]*** \\
& \vspace{2pt} \cellcolor{yellow!30}\shortstack[l]{Somewhat acceptable | \\\;\;Totally acceptable} & \cellcolor{yellow!30}784.29; [40.39, 15230]*** \\
\midrule
\multirow{4}[0]{*}{\rotatebox[origin=c]{90}{\textbf{Controlled IVs\hspace{3pt}}}}
& \textsf{creator} (Intimate partner) & 1.46; [0.66, 3.26] \\
& \cellcolor{yellow!30} \textsf{action} (Sport) & \cellcolor{yellow!30}47.55; [11.08, 204.08]*** \\
& \cellcolor{yellow!30} \textsf{action} (Saying something) & \cellcolor{yellow!30}9.96; [2.25, 44.08]** \\
& \cellcolor{yellow!30} \textsf{intent} (Entertainment) & \cellcolor{yellow!30}13.88; [3, 64.17]*** \\
& \cellcolor{yellow!30} \textsf{intent} (Sexual pleasure) & \cellcolor{yellow!30}21.13; [4.57, 97.62]*** \\
\midrule
\multirow{4}[0]{*}{\rotatebox[origin=c]{90}{\textbf{\shortstack{Uncontrolled\\IVs}}}}
& \cellcolor{yellow!30} Gender (man) & \cellcolor{yellow!30} 2.54; [1.46, 4.41]*** \\
& \cellcolor{yellow!30} \gii \& NDII attitudes (Unacceptable) & \cellcolor{yellow!30}0.19; [0.04, 0.81]* \\
& \cellcolor{yellow!30} SCS-R2 & \cellcolor{yellow!30}0.51; [0.37, 0.71]*** \\
& SCS-R4 & 1.13; [0.86, 1.48] \\
\midrule
\multirow{4}[0]{*}{\rotatebox[origin=c]{90}{\textbf{Interaction Terms\hspace{24pt}}}}
& \shortstack[l]{ \textsf{creator} (Intimate partner) \& \\\;\textsf{intent} (Entertainment)} &2.62; [0.98, 7.01] \\
& \vspace{2pt} \cellcolor{yellow!30} \shortstack[l]{\textsf{creator} (Intimate partner) \& \\\;\textsf{intent} (Sexual pleasure)} &\cellcolor{yellow!30} 3.64; [1.33, 9.96]** \\
& \vspace{2pt} \shortstack[l]{\textsf{action} (Sport) \& \\\;\textsf{intent} (Entertainment)} & 0.72; [0.14, 3.54] \\
& \vspace{2pt} \shortstack[l]{\textsf{action} (Saying something) \& \\\; \textsf{intent} (Entertainment)} & 1.70; [0.33, 8.82] \\
& \vspace{2pt} \cellcolor{yellow!30} \shortstack[l]{\textsf{action} (Sport) \& \\\;\textsf{intent} (Sexual pleasure)}&
\cellcolor{yellow!30}0.08; [0.02, 0.4]** \\
& \vspace{2pt} \cellcolor{yellow!30} \shortstack[l]{\textsf{action} (Saying something) \& \\\; \textsf{intent} (Sexual pleasure)} & \cellcolor{yellow!30} 0.19; [0.04, 1]* \\
\bottomrule
\end{tabular}
\caption{Results from a single regression exploring the relationship between the acceptability of \textsf{creation} (first row, intercepts), contextual factors (second row, controlled IVs), personal factors (third row, uncontrolled IVs), and interactions between \textsf{intent} and \textsf{creator} or \textsf{action} (fourth row, interaction terms). Reference levels: \textsf{creator} (stranger), \textsf{action} (sexual act), \textsf{intent} (harm), gender (woman), \gii \& NDII attitudes (acceptable). Significance of OR: $p<0.05$ = \colorbox{yellow!30}{*}, $p<0.01$ = \colorbox{yellow!30}{**}, and $p<0.001$ = \colorbox{yellow!30}{***}.}
\label{tab:contextual-interaction-men-women}
\end{table}

\ifnum\confversion=0
\vspace{2em}
\section{Additional Figures}
\label{ap:unacceptable-code-heatmaps}
\label{ap:acceptable-code-heatmaps}
Odds ratios and confidence intervals for the \textsf{creation}, \textsf{private\_sharing}, \textsf{public\_sharing}, \textsf{resharing}, and \textsf{seeking\_out} models are visualized in Figure~\ref{fig:ORGraphs}. 
Figure~\ref{fig:GIIHeatmaps-full} depicts heatmaps comparing acceptability involving AIG-NCII and GII created in an intimate partnership. Figure~\ref{fig:acceptable-code-distribution} and Figure~\ref{fig:unacceptable-code-distribution} visualize the distribution of codes for justifications of acceptable and unacceptable \textsf{creation}, respectively, across vignettes and \textsf{actions}.

\label{ap:or-graphs}
\begin{figure*}
  \centering
  \includegraphics[width=1\textwidth]{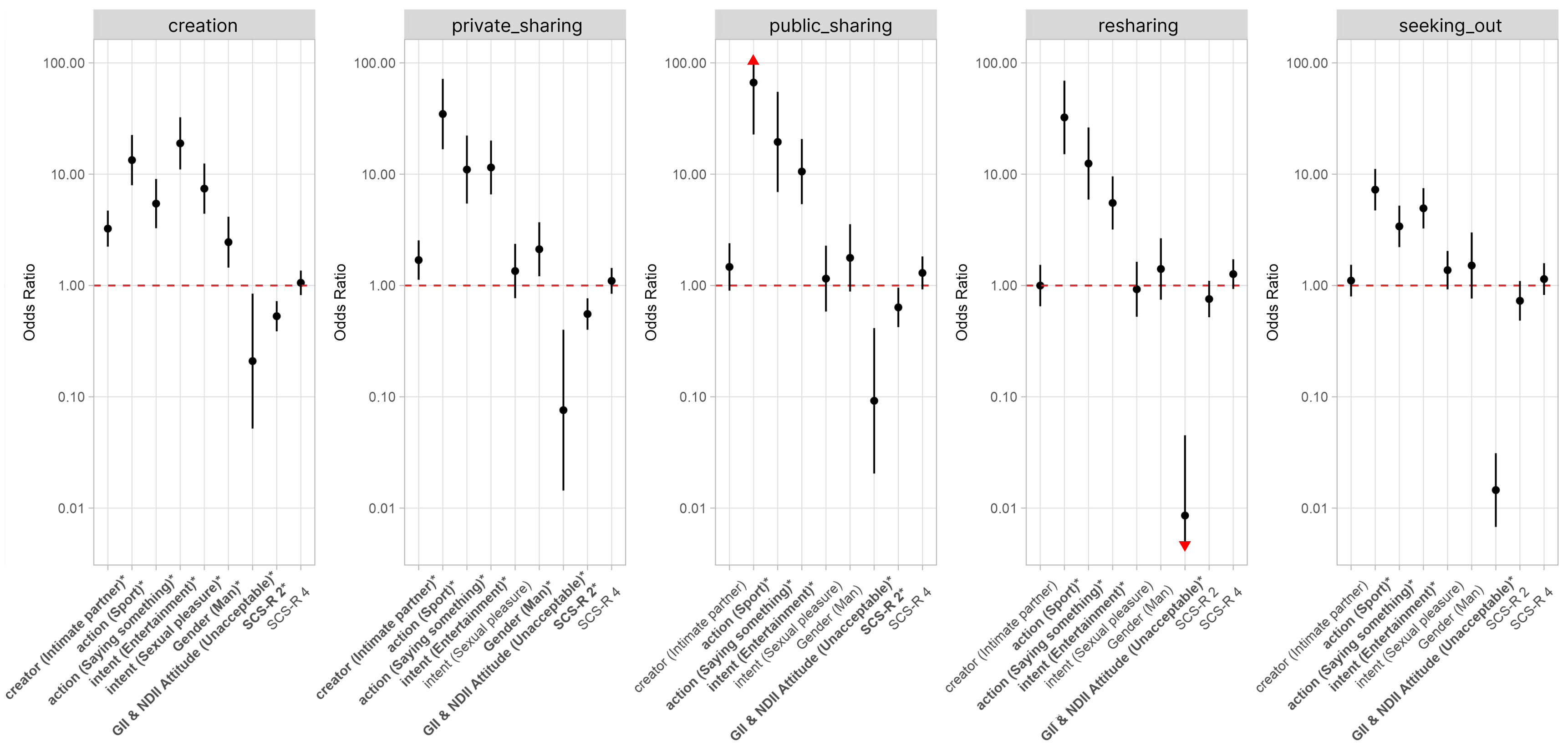}
  \caption{Odds ratios and confidence intervals for the controlled and uncontrolled IVs in the models for \textsf{creation}, \textsf{private\_sharing}, \textsf{public\_sharing}, \textsf{resharing}, and \textsf{seeking\_out}. Statistically significant IVs ($p<0.05$) are in bold with an asterisk. Y-axis is on a logarithmic scale and trims are indicated by red triangles.}
  \label{fig:ORGraphs}
\end{figure*}

\label{ap:GIIHeatmaps-full}
\begin{figure*}[]
  \centering
  \includegraphics[width=1\textwidth]{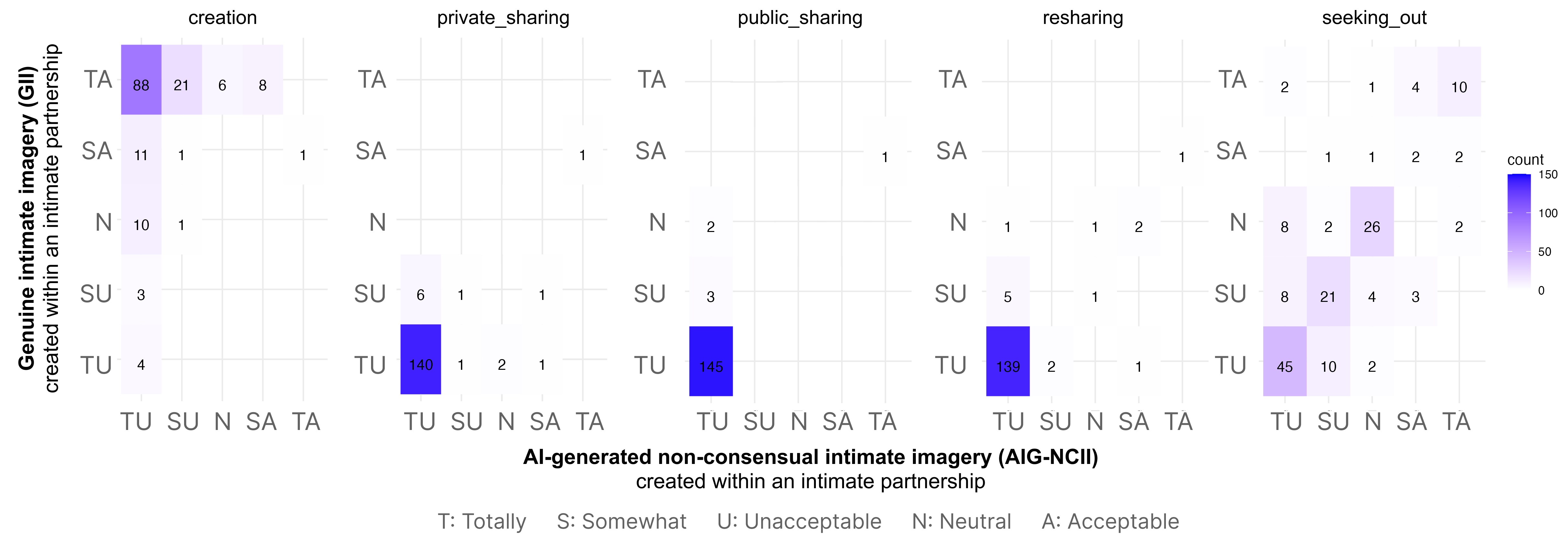}
  \caption{Heatmap of acceptability for \textsf{creation} and \textsf{seeking\_out} when the 
  \textsf{action} is performing a sexual act. Darker indicates more respondents.}
  \label{fig:GIIHeatmaps-full}
\end{figure*}

\begin{figure*}
  \centering
  \includegraphics[width=1\textwidth]{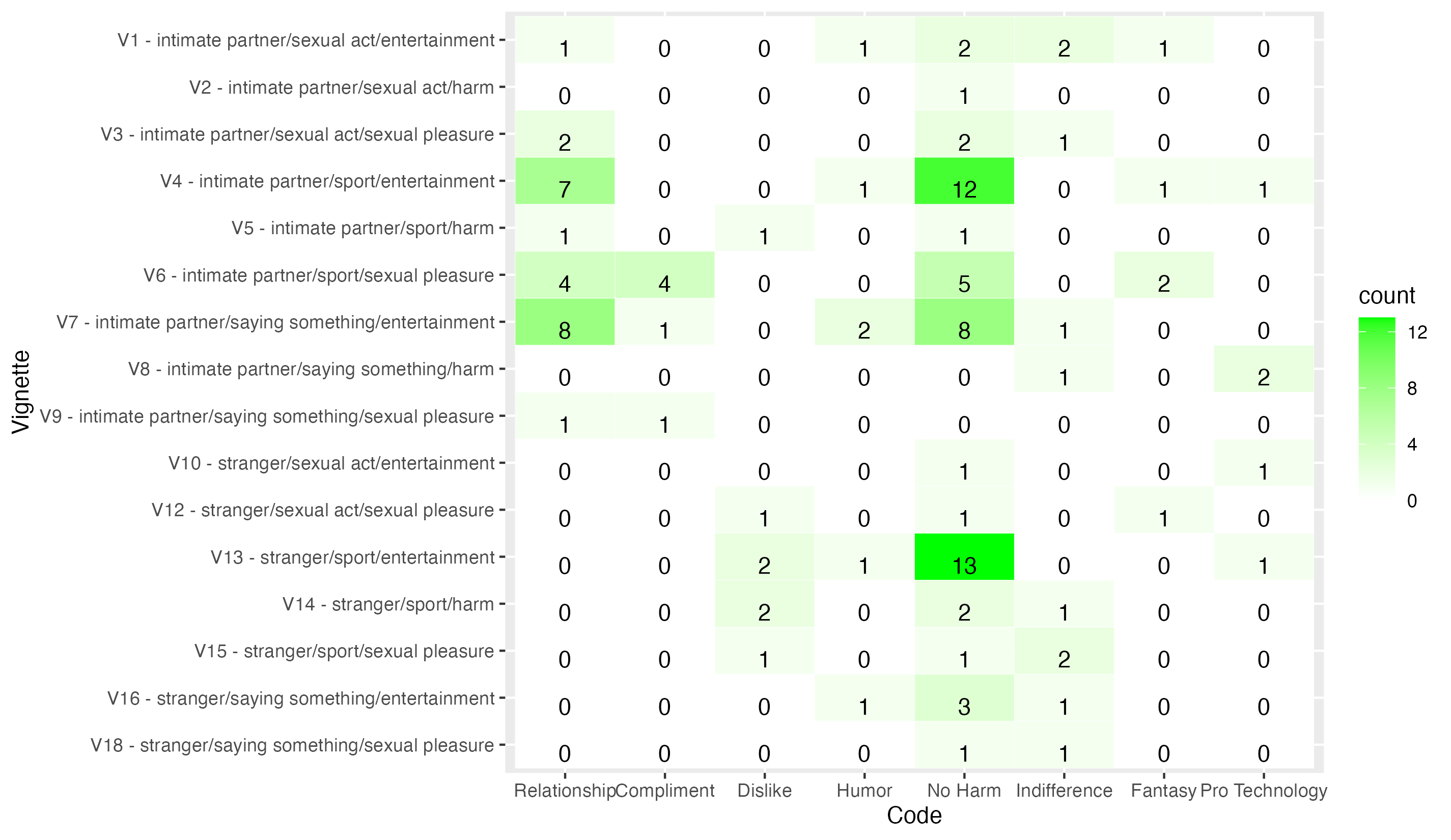}
  \includegraphics[width=0.85\textwidth]{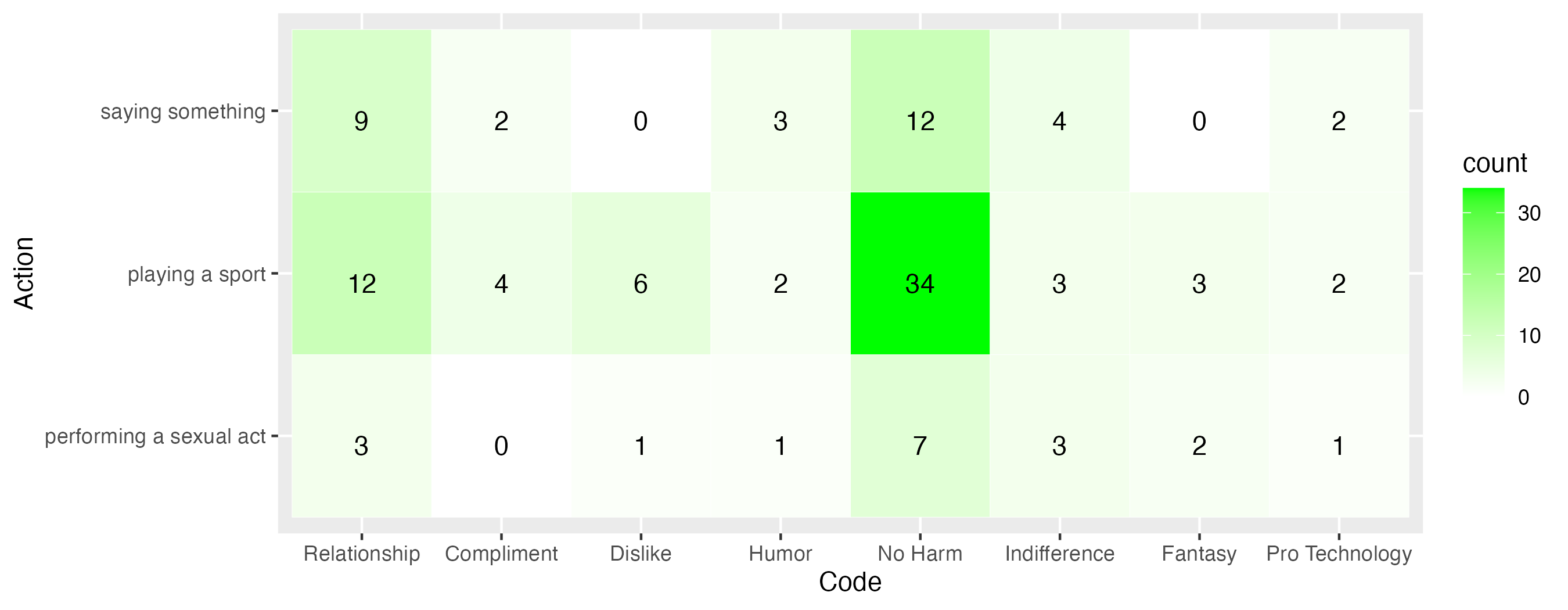}
  \caption{Heatmaps of the frequency of codes for justifications of acceptable \textsf{creation}. The top heatmap shows occurrences for each vignette and the bottom heatmap shows occurrences across \textsf{actions}. Darker green indicates a higher count of occurrences in the qualitative data.}
  \label{fig:acceptable-code-distribution}
\end{figure*}

\begin{figure*}
  \centering
  \includegraphics[width=1\textwidth]{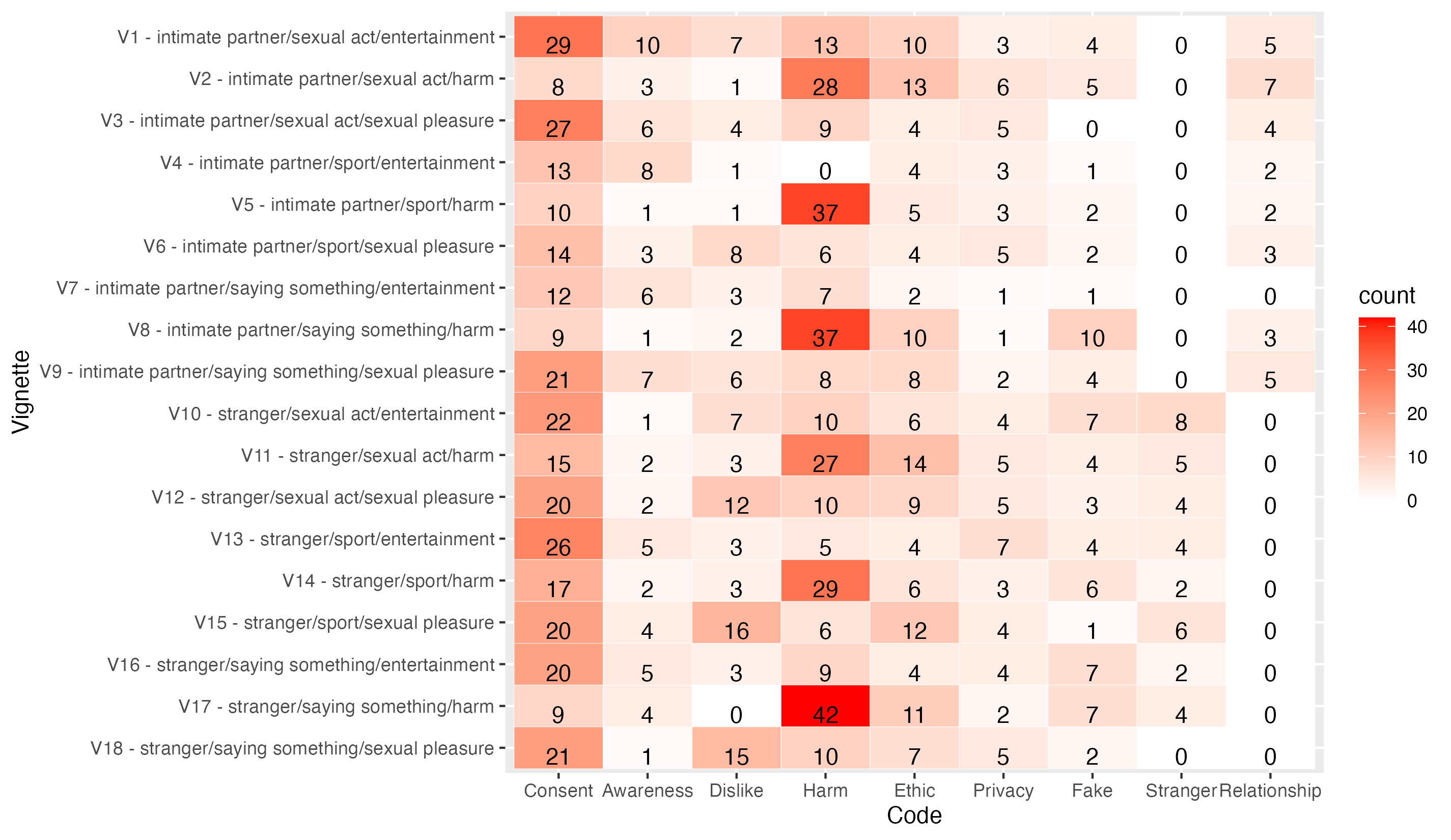}
  \includegraphics[width=0.85\textwidth]{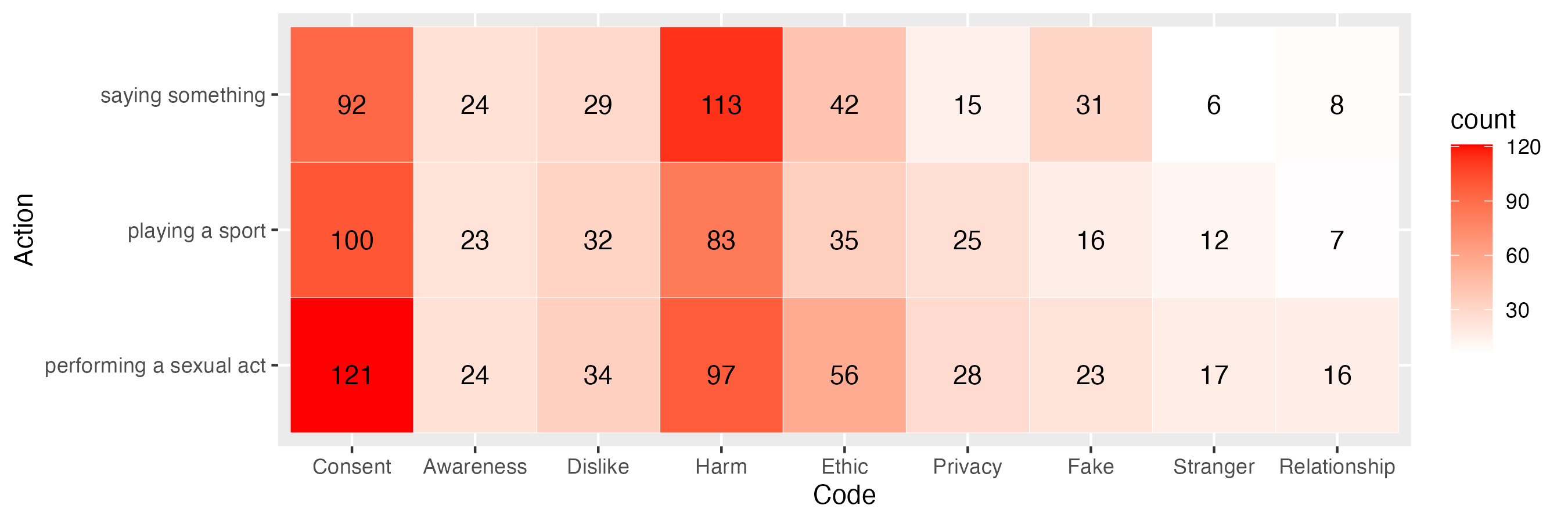}
  \caption{Heatmaps of the frequency of codes for justifications of unacceptable \textsf{creation}. The top heatmap shows occurrences for each vignette and the bottom heatmap shows occurrences across \textsf{actions}. Darker red indicates a higher count of occurrences in the qualitative data.}
  \label{fig:unacceptable-code-distribution}
\end{figure*}
\fi

\end{document}